\documentclass[12pt]{iopart}
\usepackage{graphicx}
\usepackage{epsfig,color}
\usepackage{hyperref}

\expandafter\let\csname equation*\endcsname\relax
\expandafter\let\csname endequation*\endcsname\relax
\usepackage{amsmath}
\usepackage{amsfonts}
\usepackage{amssymb}
\usepackage{cite}

\newcommand{\ket}[1]{|{#1}\rangle}
\newcommand{\bra}[1]{\langle{#1}|}

\newcommand{\bracket}[2]{\langle#1|#2\rangle}

\begin{document}

\title{Building Projected Entangled Pair States with a Local Gauge Symmetry}

\date{\today}

\author{Erez Zohar and Michele Burrello}
\address{Max-Planck-Institut f\"ur Quantenoptik, Hans-Kopfermann-Stra\ss e 1, 85748 Garching, Germany}

\begin{abstract}
Tensor network states, and in particular projected entangled pair states (PEPS), suggest an innovative approach for the study of lattice gauge theories, both from a pure theoretic point of view, and as a tool for the analysis of the recent proposals for quantum simulations of lattice gauge theories. In this paper we present a framework for describing locally gauge invariant states on lattices using PEPS. The PEPS constructed hereby shall include both bosonic and fermionic states, suitable for all combinations of matter and gauge fields in lattice gauge theories defined by either finite or compact Lie groups.
\end{abstract}

\maketitle

\section{Introduction}
The importance of the standard model of particle physics, and within it, of gauge theories, could not be overestimated. Nevertheless, a very important part of it, Quantum Chromodynamics (QCD), remains partially unresolved, due to its highly non-perturbative nature.
One of the ways to tackle the non-perturbative physics of the strong interactions described by QCD is lattice gauge theories \cite{Wilson,KogutSusskind,KogutLattice,Kogut1983}, which have been very successful in the study of a broad range of phenomena \cite{Bali1992,Philipsen1998,Knechtli1998,Chernodub2002,Bali2005,Durr2008,Pepe2009,Fukushima2011}. The most traditional methods of investigation of lattice gauge theories are based on Monte-Carlo calculations in a Euclidean space-time and, in spite of their unquestionable efficiency to achieve many tasks, they present some limitations in specific cases. For example, the computationally hard sign-problem \cite{Troyer2005} hinders calculations in fermionic systems with a finite chemical potential, a situation which is important for several QCD phases, such as color superconductivity and quark-gluon plasma \cite{Fukushima2011,McLerran1986}. Another traditionally difficult task for Monte-Carlo techniques is the implementation of the real-time evolution in Minkowski spacetime: despite the development of new techniques \cite{Hebenstreit2013a,Hebenstreit2013} which are fruitful for various calculations, real-time evolution is still challenging in the most general situations.

Recently, alternative approaches to examine lattice gauge theories have been considered and there have been several proposals for quantum simulations \cite{Feynman1982,CiracZoller}  of lattice gauge theories \cite{Zohar2011,Zohar2012,Banerjee2012,Zohar2013,Tagliacozzo2013,NA,Rishon2012,TagliacozzoNA,Topological,AngMom,ZollerIons,SQC,Wiese2013,Dissipation,
 Marcos2014,Wiese2014,Zohar2015a,Mezzacapo2015,Pepe2015,LaFlamme2015}, in which such models are mapped to AMO systems (mostly ultracold atoms in optical lattices \cite{Lewenstein2012}), which are highly controllable in the laboratory and hence serve as \emph{analog- or digital-quantum-computers} specially tailored to these  problems.

 Besides the quantum simulation proposals, new ways for classical simulations of lattice gauge theories have been developed and utilized, based on tensor network states, in particular matrix product states (MPS) and projected entangled pair states (PEPS) \cite{Verstraete2004,Schuch2010}; see, for example,
   \cite{Byrnes2002,Banuls2013,Banuls2013a,Rico2013,Tagliacozzo2014,Kuhn2014,Haegeman2014,Silvi2014,Buyens2014,Saito2014,Banuls2015,Pichler2015,Kuhn2015,Zohar2015b,Buyens2015}.
   These works are of great importance, as they suggest new ways to overcome the problems of the conventional Monte-Carlo calculations and possibly to observe and understand new physical phenomena. They are also closely related to quantum simulations, as quantum simulators usually require some truncation of the Hilbert space of the gauge theories \cite{Zohar2015}, and tensor network studies may help shedding light on the quantitative differences between the quantum simulators and the full models.

The study of lattice gauge theories within the tensor network framework takes its steps from the techniques developed to describe quantum states with symmetry groups, which allow us to directly encode these symmetries in the properties of the adopted tensors, especially in the case of translational invariant states. Several works analyzed in detail the construction of both one-dimensional states with global symmetries \cite{Singh2010,Singh2010b,Schuch2011,Weichselbaum2012} and two-dimensional PEPS with spatial symmetries and global gauge symmetries acting on inner degrees of freedom \cite{Perezgarcia2010,Singh2011,Schuch2011}. In the description of gauge theories a further step is required: physical gauge degrees of freedom must be included to gauge the symmetries from global to local.

This last step has been undertaken for two-dimensional tensor networks in the recent works \cite{Tagliacozzo2014,Haegeman2014,Zohar2015b}. In particular, Tagliacozzo {\it et al.} \cite{Tagliacozzo2014} introduced a formalism for building bosonic PEPS describing pure lattice gauge theories with arbitrary groups and adopted a truncation scheme for the representations which we will extend in this work to include matter and fermionic PEPS as well; Haegeman {\it et al.} \cite{Haegeman2014} developed, instead, a general, conceptual formalism to build PEPS which describe both gauge fields  and bosonic matter through a gauging map of injective PEPS. Finally, in \cite{Zohar2015b}, the authors presented an explicit analytical and numerical construction for the specific case of a truncated $U(1)$ gauge-invariant state including both fermionic matter and bosonic gauge fields.

In this work, we develop a more general framework for the construction of PEPS with local gauge invariance
 for lattice gauge theories whose gauge group $G$ is either a compact Lie, or a finite group. We aim at unifying all the elements required for complete descriptions of lattice gauge theories: in particular we focus on gauge invariant states which include both matter and gauge fields and we propose a constructive approach to define tensor network states where the matter can be either bosonic or fermionic and is associated to an arbitrary representation of the gauge group. A special attention will indeed be devoted to the construction of fermionic PEPS because of their more immediate relevance in the study of models which are closer to the usual particle physics theories.

Throughout this paper we adopt the mathematical stucture discussed in \cite{Zohar2015} to describe the physical components of the PEPS and, in particular, the Hilbert spaces used to describe the matter and the gauge field degrees of freedom. We will show that such description, based on the representations of the gauge group, can be naturally extended to the virtual states constituting the links of the tensor network. Furthermore, to make this work self-contained, we will summarize the main elements of the construction presented in \cite{Zohar2015} without assuming  that the reader is acquainted with PEPS or lattice gauge theories; the required PEPS details are reviewed and constructed along the paper, as well as their relations with lattice gauge theory.

\section{Mathematical framework for the description of lattice gauge theories} \label{sec:math}

The purpose of this paper is to build, with a tensor network approach, physical states which fulfill a local gauge invariance under a gauge group $G$.
These states describe many-body systems living in a two-dimensional square lattice and constitute paradigmatic or variational states for the study of lattice gauge theories. Following the standard formulation of lattice gauge theories, the vertices of the square lattice host the matter degrees of freedom, whereas its links host the degrees of freedom associated to the gauge fields.

The requirement of the gauge symmetry must be translated to the invariance under specific operators that describe the effect of the group $G$ on both the matter particles and the gauge fields mediating among them. Matter particles and gauge fields, though, are described by different Hilbert spaces and obey different transformation rules.
The matter particles require, in general, the introduction of an internal degree of freedom (the ``color'' in high-energy models) over which the group elements can act. The gauge field degrees of freedom, instead, are associated to group elements since they come into play when considering the transformation that a particle undergoes while moving between neighboring vertices.

Before turning into the PEPS construction, in this Section we shall  briefly review some basic notation and group theoretic properties required for the description of both matter particles and gauge fields. We will follow, in particular, the construction of \cite{Zohar2015}.

Consider a group $G$, which may be either  a finite or a Lie group. We denote its elements by $g \in G$, and, since in general $G$ may be non-Abelian, right and left group transformations have to be defined separately. We denote the quantum unitary operators corresponding to these right and left transformations $\Theta_g$ and $\tilde \Theta_g$ respectively. In order to represent their action on the matter degrees of freedom by matrices, a suitable Hilbert space must be defined.

This Hilbert space is spanned by the states $\left|jm\right\rangle$.
First, $j$ denotes the \emph{representation} of the group under which the state is transformed: the transformations $\Theta_g$
and $\tilde \Theta_g$ are block diagonal in terms of the representation $j$ - a direct sum
of matrices with dimensions $\dim\left(j\right)$ acting on the $j$ subspaces. For $SU(2)$, for example, the
 representations are identified with the total angular momentum of a state, and the states are eigenvalues of the quadratic Casimir operator $\mathbf{J}^2$,
\begin{equation}
\mathbf{J}^2 \left|jm\right\rangle = j\left(j+1\right)\left|jm\right\rangle
\end{equation}
a similar operator may be defined for other Lie groups (such as $SU(N)$), and in any case one may define a general operator of the form
\begin{equation}
\underset{j}{\sum}f\left(j\right)\left|jm\right\rangle\left\langle jm\right|
\end{equation}
whose eigenvalues are the representations. Here, and throughout this paper, a summation is assumed on double indices, except for those corresponding to representations.

For rank 1 groups, such as $SU(2)$, every representation appears once and $j$ may be treated as a simple index; in other groups, representations may have multiplicities $d_j$ (in $SU(N)$, for example, there are $N-1$ fundamental representations), and one has to add to $j$ another index, $i\in{1...d_j}$; in general, we shall treat $j$ as either a single index or a set
of two indices corresponding to a particular representation.

To distinguish the orthogonal states within a given representation $j$, we must define sets of group operators which are block diagonal in the representations as well - such as the angular momentum operators $J_i$ for $SU(2)$, for example. Then,
out of these operators we may select a maximal set of mutually commuting operators (which will commute with the representation as well, of course, thanks to the block-diagonal structure).
The mutual commutation allows for mutual diagonalization of these operators along with the representation, which brings us to the definition of $m$ - a set of eigenvalues of these operators.
In the case of $SU(2)$ the set of such mutually commuting operators has only one element - a single component of the angular momentum, usually taken to be $J_z$, and there, indeed, $J_z\left|jm\right\rangle = m\left|jm\right\rangle$. In $SU(3)$, on the other hand, there are two such operators, corresponding (in particle physics terms) to the isospin and the hypercharge, and thus $m$ is a set of two quantum numbers.

We define the \emph{Wigner matrices} as the unitary representation matrices of the right transformations,
\begin{equation}
\left\langle jm \right|\Theta_g\left|jn\right\rangle \equiv D^{j}_{mn}\left(g\right);
\end{equation}
then, the right transformation law under $G$ is
\begin{equation} \label{grouptransformation}
\Theta_g\left|jm\right\rangle = D^{j}_{nm}\left(g\right) \left|jn\right\rangle.
\end{equation}
All the $\dim\left(j\right)$ states of the representation $j$ are, in general, mixed by such transformations, but only among themselves.  The group element $g$ determines the amplitudes of the transformation in \eqref{grouptransformation}.

To understand the name ``right transformation'', we define a fermionic creation operator $a^{j\dagger}_m$, such that
\begin{equation}
\left|jm\right\rangle = a^{j\dagger}_m\left|0\right\rangle
\end{equation}
where $\left|0\right\rangle$ is a group singlet - invariant under any transformation - corresponding to the fermionic vacuum.
Then, the transformation law
\begin{equation} \label{righttr}
\Theta_g a^{j\dagger}_m \Theta^{\dagger}_g = a^{j\dagger}_n D^{j}_{nm}\left(g\right)
\end{equation}
simply multiplies the vector (rank-1 tensor) $a^{j\dagger}_m$ \emph{from the right} by the representation matrix corresponding to $g$, thus reproducing the right transformation law of the group states. From that, it is straightforward to define the ``left transformation''
\begin{equation} \label{lefttr}
\tilde \Theta_g a^{j\dagger}_m \tilde \Theta^{\dagger}_g = D^{j}_{mn}\left(g\right) a^{j\dagger}_n
\end{equation}
out of which one obtains
\begin{equation}
\tilde \Theta_g\left|jm\right\rangle = D^{j}_{mn}\left(g\right) \left|jn\right\rangle
\end{equation}
as well as
\begin{equation}
\left\langle jm \right|\tilde \Theta_g\left|jn\right\rangle \equiv D^{j}_{nm}\left(g\right).
\end{equation}

Note that these fermionic operators both have well defined right and left transformation laws, and thus they are not equivalent to the rishon modes present in the link model formulation of lattice gauge theories \cite{wiese1997}. The fermionic operators, in our case, will not be used for the construction of the physical electric field.

To describe the gauge field, we must instead define a matrix operator, or a rank-2 tensor under the group, $U^{j}_{mn}$, such that:
\begin{equation} \label{righttr2}
\Theta_g U^j_{mn} \Theta^{\dagger}_g = U^j_{mn'} D^j_{n'n}\left(g\right)
\end{equation}
and
\begin{equation} \label{lefttr2}
\tilde \Theta_g U^j_{mn} \tilde \Theta^{\dagger}_g = D^j_{mm'}\left(g\right) U^j_{m'n}
\end{equation}

These operators act on a different Hilbert space, which describes the physical states along the links of the square lattice and is spanned by the states $\left|jmn\right\rangle$, which we define as:
\begin{equation} \label{gaugestates}
\left|jmn\right\rangle  \equiv \sqrt{\dim(j)} \, U^j_{mn} \left|000\right\rangle,
\end{equation}
based on a trivial state $\ket{000}$. This results in their transformation rules
\begin{equation}
\Theta_g \left|jmn\right\rangle = D^j_{n'n}\left(g\right)\left|jmn'\right\rangle
\end{equation}
and
\begin{equation}
\tilde \Theta_g \left|jmn\right\rangle = D^j_{mm'}\left(g\right)\left|jm'n\right\rangle.
\end{equation}
These transformations act separately on the $m,n$ numbers, which define the \emph{left} and \emph{right} degrees of freedom: the states $\left|jmn\right\rangle$ are what we call \emph{representation states}, or elements of the \emph{representation basis} of the gauge field, and may be seen as
\begin{equation}
\left|jmn\right\rangle = \left|jm\right\rangle \otimes \left|jn\right\rangle
\end{equation}
- i.e., this Hilbert space consists of two copies of the previous Hilbert space with the restriction, however, that the two copies share the same representation.
This equivalence, however, is only in the level of group transformation laws, and does not mean that the $U$ operators are created out of two $a$ operators representing right and left modes; as mentioned before, the fermionic $a$ operators are not intended to play the role of ``left'' or ``right'' particles. This sets a difference from the link model rishon formulation \cite{wiese1997} and their tensor network simulations \cite{Silvi2014}. Differently from the link model formulation, in our construction the right and left part are not separate ingredients and they do not generate any multiplicity. Furthermore the truncation scheme we adopt will be different.

Then it is straightforward, using the Clebsch-Gordan series, to expand the $U^{j}_{mn}$ operators in the representation basis \cite{Zohar2015},
\begin{equation}
 U_{mm'}^{j} = \underset{J,K}{\sum} \sqrt{\frac{\text{dim}\left(J\right)}{\text{dim}\left(K\right)}}  \bracket{JMjm}{KN} \left\langle KN' | JM'jm' \right\rangle
 \ket{KNN'}\bra{JMM'}. \label{Udef}
\end{equation}
One can verify that $U_{mm'}^{j}$ is in general a unitary operator.

For the practical purpose of simulating gauge fields with a finite Hilbert space, it is usually necessary to restrict the set of representations that can appear on the link states. The previous decomposition of the operator $U_{mm'}^{j}$ offer a possible and consistent way of applying this restriction: the sum on the right hand side of \eqref{Udef} can be truncated in a way which maintains unaltered all the previous the transformation rules. This is achieved just by considering only a small set of representations which are connected via non-vanishing Clebsch-Gordan coefficients \cite{Zohar2015}, and discarding all the other terms.

An alternative interpretation of $U^{j}_{mn}$ is given by the fact that this operator plays the role of the connection entering the kinetic term in the Hamiltonian of lattice gauge theories \cite{KogutSusskind} which assumes the form:
\begin{equation} \label{tunneling}
  \psi^{j_p \dag}_{m}\left({\mathbf{x}}\right) U^{j_p}_{mn} ( \mathbf{x},\hat{k})   \psi^{j_p}_n( \mathbf{x}+\hat{k}) + {\rm H. c.}
\end{equation}
where $\psi^{j_p}_{m}({\mathbf{x}})$ is the annihilation operator of the form $a^{j_p}_m$ of a physical matter fermion in the lattice vertex $\mathbf{x}$, with $j_p$ the group representation associated to the single particles of matter. $\hat{k}$ is a unit vector specifying a direction of the lattice and the gauge field degrees of freedom on the link between $\mathbf{x}$ and $\mathbf{x}+\hat{k}$ are affected by the operator $U^{j_p}_{mn} ( \mathbf{x},\hat{k})$.
Hence we can consider $U^{j}_{mn}$ as a quantum operator whose spectrum is given by the Wigner matrix elements,
\begin{equation}
U^{j}_{mn} = \int dg D^{j}_{mn}\left(g\right) \left|g\right\rangle\left\langle g\right|.
\end{equation}
since it corresponds to the the transformation that matter particles undergo when hopping from one site to another; therefore $U^{j_p}_{mn}$ must be diagonal in the \emph{group element basis}, whose elements are states corresponding to elements of the group - $\left|g\right\rangle$.

The group element basis and the representation basis are connected through the relation
\begin{equation}
\left\langle g | jmn \right\rangle = \sqrt{\frac{\dim\left(g\right)}{\left|G\right|}}D^{j}_{mn}\left(g\right)
\label{FTrans}
\end{equation}
where $\left|G\right|$ is the order of the group. In that sense, the Wigner matrices may be seen as wavefunctions of the representation states in a coordinate space parametrized by the group elements.

It is important to notice that the right and left transformation rules in Eqs. (\ref{righttr},\ref{righttr2}) and (\ref{lefttr},\ref{lefttr2}) respectively enforce the gauge invariance of the tunneling term \eqref{tunneling}, once we consider a suitable convention for the orientation of the link.

Concerning the gauge field degrees of freedom, the choice of the representation or the group basis reflects the different interpretations in terms of electric field and vector potential. Electric energy terms in the Kogut-Susskind Hamiltonian for $SU(N)$ \cite{KogutSusskind} are given in terms of the representations - the square of the electric field is identified with the quadratic Casimir operator. Thus, representation states may be also referred to as \emph{electric states}, and thus a Hilbert space which is truncated in terms of this basis is suitable for electric states or states in phases with electric confinement, where the electric flux is concentrated. On the other hand, the magnetic energy is given by traces of the sum of $U^{j}_{mn}$ products along a plaquette, and thus group element stats may be referred to as \emph{magnetic states}, since they are the eigenvalues of these.

A better intuition for this formalism may be given if the group $G$ is a Lie group. Then, one may use group parameters and generators to parameterize the quantum transformation operators.

We can consider in further detail the case in which $G$ is a compact Lie group. There, the transformation operators $\Theta_g$ may be written in terms of parameters and generators:
\begin{equation} \label{Lieright}
\Theta_g = e^{i \mathbf{q}_g \cdot \mathbf{R}}
\end{equation}
and
\begin{equation} \label{Lieleft}
\tilde \Theta_g = e^{i \mathbf{q}_g \cdot \mathbf{L}},
\end{equation}
where $\mathbf{q}_g$ is the vector of group parameters for $g$, and $\mathbf{L,R}$ are the generators of transformations for the left and right transformations, satisfying the algebra
\begin{equation}
\begin{aligned}
& \left[L_a,L_b\right]=-if_{abc}L_c \\  & \left[R_a,R_b\right]=if_{abc}R_c \\ & \left[L_a,R_b\right]=0 \\ & \left[U^{j}_{mn},L_a\right]=T^{j}_{mm'}U^{j}_{m'n}
\\ & \left[U^{j}_{mn},R_a\right]=U^{j}_{mn'}T^{j}_{n'n,}
\end{aligned}
\end{equation}
where $f_{abc}$ are the structure constants of $G$, and $T^{j}$ are the $j$-representation matrices of the generators of $G$. $R_a,L_a$ are then identified as the right and left electric fields on a link.

One may define the quadratic Casimir operator, $\mathbf{J}^2 = \mathbf{E}^2 \equiv L_aL_a = R_aR_a$, and then
$\mathbf{E}^2\left|jmn\right\rangle = C_2\left(j\right)\left|jmn\right\rangle$, where $C_2$  is a function of the representation only, representing the square of the electric field.

 For $G=SU(2)$, for example, $C_2\left(j\right) = j\left(j+1\right)$. $m$ and $n$ represent sets of eigenvalues of some subset of $L_a,R_a$ operators, mutually commuting. In this case these are the eigenvalues of the $z$ components of the angular momentum, i.e. $L_z\left|jmn\right\rangle = m\left|jmn\right\rangle$ and $R_z\left|jmn\right\rangle=n\left|jmn\right\rangle$: one has two sets of angular momenta, left and right one, corresponding to the angular momentum of a rigid body, measured in two different systems of principal axes: the space inertial frame and the body rotating frame; the total angular momentum of both frames has to be equal since it is a scalar under rotations.
 In fact, $\left|jmn\right\rangle$ are simply eigenstates of the free isotropic rigid rotator \cite{KogutSusskind,Landau1981}, described by the Hamiltonian
\begin{equation}
H = \frac{\mathbf{J}^2}{2I}
\end{equation}
the eigenfunctions of this Hamiltonian in coordinate space - the space of Euler angles $\alpha,\beta,\gamma$ were found by Wigner (up to some conventions) to be
\begin{equation}
\left\langle \alpha,\beta,\gamma | jmn \right\rangle = \sqrt{\frac{2j+1}{8\pi^2}}D^{j}_{mn}\left(\alpha,\beta,\gamma\right)
\end{equation}
- which is simply the $SU(2)$ case of eq. (\ref{FTrans}).

\section{PEPS with a global symmetry}

\begin{figure}
  \centering
  \includegraphics[width=\textwidth]{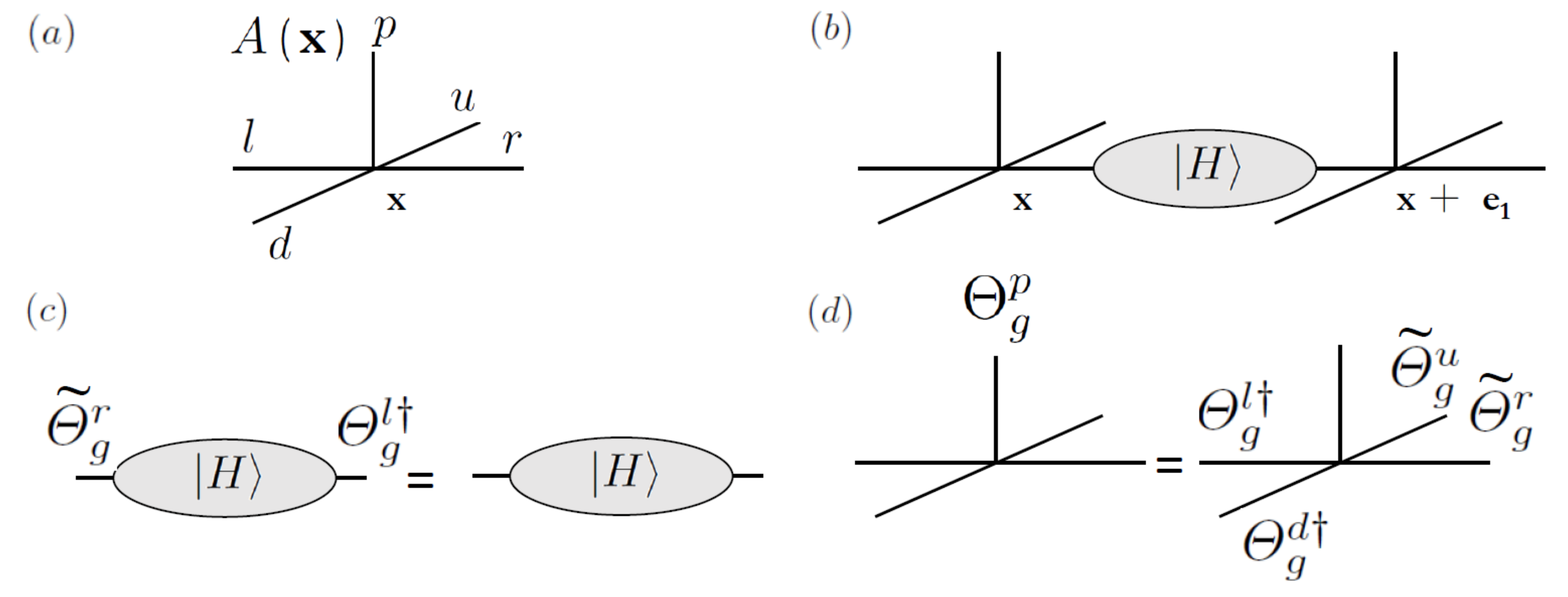}\\
  \caption{The $A$ tensor and its symmetries. (a) The vertex tensor, with the physical state $p$ and the virtual states $l,r,u,d$, as defined
  in (\ref{phindef}). (b) Projection to a maximally entangled state on the bond. (c) The symmetry of the fiducial state - a virtual Gauss law
  (\ref{eqsym1}). (d) Invariance of the projectors under $G$ (\ref{eqinv}).}\label{fig1}
\end{figure}

Before considering local gauge symmetries, which are at the main focus of this paper, let us recall the structure of PEPS in the case of physical states invariant under the action of a symmetry group (see, for example, \cite{Perezgarcia2010,Singh2011}). In our case, this symmetry group will give rise to a \emph{global gauge symmetry}, acting, in general, on an inner degree of freedom of the physical matter particles.

We consider a two dimensional (spatial) square lattice, and  begin our construction by associating a \emph{fiducial state} to each of its sites $\mathbf{x}$. These local fiducial states are obtained by the use of tensors $A^{p}_{lrud}$, and assume the form:
\begin{equation}
\left|A\left(\mathbf{x}\right)\right\rangle = A^{p}_{lrud}\left|p\right\rangle\left|l,r,u,d\right\rangle
\label{phindef}
\end{equation}
where summation on the indices $p,l,r,u,d$ is implied. The $\left|p\right\rangle$ state belongs to the \emph{physical} Hilbert space $\mathcal{H}_p$ on the vertex $\mathbf{x}$, with dimension $d$; the states $\left|l,r,u,d\right\rangle$ belong  to four \emph{virtual} identical Hilbert spaces $\otimes \mathcal{H}_v$. Each of them has dimension $D$ (called the \emph{Bond Dimension}) and it is associated to one of the edges of the links departing from the vertex $\mathbf{x}$ (see figure \ref{fig1}a).
In order to obtain a physical many-body state, several local fiducial states must be connected by suitable projectors. To this purpose we must first define a global fiducial state as the tensor product of the previous local states on the vertices of the lattice. In particular, we will impose to have  translational invariance, which is implemented by choosing the same tensor $A$ in all the positions (this condition will be slightly altered Sec. \ref{sec:fermions} to include a staggering in the fermionic systems to account for positive and negative charges). The role of possible boundaries is  neglected here for the sake of simplicity although it can be handled with standard tensor network techniques (see, for example, \cite{Zohar2015b}, for the specific case of the U(1) gauge symmetry). This choice of building a translational invariant state reflects the translational symmetry that is customarily considered in the study of lattice gauge theories without background static charges. This results in the product state:
\begin{equation}
\left|\Phi\left(A\right)\right\rangle = \underset{\mathbf{x}}{\bigotimes}\left|A\left(\mathbf{x}\right)\right\rangle.
\end{equation}
We also define the following maximally entangled states along the bonds (see figure \ref{fig1}b),
\begin{equation} \label{Hstate}
\left|H\left(\mathbf{x},\mathbf{x+\hat e}_1\right)\right\rangle = \frac{1}{\sqrt{D}}\underset{i \in \mathcal{H}_v}{\sum}\left|r\left(\mathbf{x}\right)=i
,l\left(\mathbf{x+\hat e}_1\right)=i\right\rangle
\end{equation} and
\begin{equation} \label{Vstate}
\left|V\left(\mathbf{x},\mathbf{x+\hat e}_2\right)\right\rangle = \frac{1}{\sqrt{D}}\underset{i \in \mathcal{H}_v}{\sum}\left|u\left(\mathbf{x}\right)=i
,d\left(\mathbf{x+\hat e}_2\right)=i\right\rangle
\end{equation}
for the horizontal and vertical links (bonds) of the lattice respectively \footnote{One may, of course, replace these by other, similar entangled bond states.}.
The product state of all links (bonds) shall be denoted as
$\left|\left\{HV\right\}\right\rangle = \underset{\mathbf{x}}{\bigotimes}
\left|H\left(\mathbf{x},\mathbf{x+\hat e}_1\right)\right\rangle
\left|V\left(\mathbf{x},\mathbf{x+\hat e}_2\right)\right\rangle$.
Eventually, one projects the fiducial state $\ket{\Phi(A)}$ onto the entangled bond states $\ket{HV}$, thus factoring the virtual states out (and hence their name). This allows us to correlate the physical states in different sites and to create a many-body
physical state, in which the indices of the $A^p_{lrud}$ tensors are properly contracted:
\begin{equation}
\left|\psi\right\rangle = \left\langle \left\{ HV \right\} | \Phi \left(A\right) \right\rangle.
\end{equation}

Suppose we wish our state to be globally invariant under some transformation group $G$.
We then parametrize all our states in terms of their transformation properties under $G$: i.e., by an irreducible representation $j$ and a set of eigenvalues $m$  of mutually commuting operators (identifiers within the representation). For example, the physical state will be denoted as $\left|j_p m_p\right\rangle$ .

Each such $\left|jm\right\rangle$ state may undergo either \emph{right} or \emph{left group transformations}, parameterized by the group elements $g$, which are represented by the unitary Wigner matrices
\begin{equation}
D^{j}_{mn}\left(g\right) = \bar{D}^{j}_{nm}\left(g ^{-1}\right).
\end{equation}
as explained in section \ref{sec:math}.

The physical and virtual states may belong to any representation. However, if we further impose that the two virtual states lying on the edges of a single link belong to the same representation, we get that the entangled states $\left|H\right\rangle,\left|V\right\rangle$ are invariant under the above group transformations (see figure \ref{fig1}c). For example,
\begin{equation}
\begin{aligned}
& \tilde \varTheta^r_g \varTheta^{l \dagger}_g\left|H\right\rangle = \frac{1}{\sqrt{D}}\tilde \varTheta^r_g \left|j m\right\rangle_r \varTheta^{l \dagger}_g\left|j m\right\rangle_l= \\
& \frac{1}{\sqrt{D}} D^{j}_{mm'}\left(g\right) \left|j m'\right\rangle_r D^{j}_{m''m}\left(g^{-1}\right)\left|j m''\right\rangle_l=
\left|H\right\rangle.
\label{eqinv}
\end{aligned}
\end{equation}
Hereafter we will adopt the notation $\varTheta_g$ for gauge operators acting on the virtual states, and $\Theta_g$ for operators acting on the physical level. In Eq. \eqref{eqinv} the indices $l$ and $r$ simply refer to the two states on the edges of the virtual bond \footnote{The superscripts of the gauge operators $\Theta$ are added just for the sake of clarity and,  differently from the indices in \eqref{phindef}, they are not indices to be summed.} and we exploited the unitarity of the Wigner matrices $D$.  Similarly, for vertical bonds, $\tilde\varTheta^{u}_g \varTheta^{d \dagger}_g\left|V\right\rangle = \left|V\right\rangle$.

We wish our state $\left|\psi\right\rangle$ to be \emph{globally} invariant under $G$. This means that the state is invariant under the global transformation operator $\Theta_g = \underset{\mathbf{x}}{\bigotimes} \Theta_g^p\left(\mathbf{x}\right)$ \footnote{The state is considered as invariant also for equality up to a global phase, but we shall ignore this possibility as this corresponds, in gauge theory terms, to the presence of static charges.}:
\begin{equation}
\Theta_g \left|\psi\right\rangle =  \left|\psi\right\rangle.
\end{equation}
This is achieved if the local fiducial states $\left|A\left(\mathbf{x}\right)\right\rangle$ satisfy
\begin{equation}
\Theta^{p}_g\left(\mathbf{x}\right) \left|A\left(\mathbf{x}\right)\right\rangle =
\varTheta^{l \dagger}_g\left(\mathbf{x}\right) \tilde \varTheta^{r}_g\left(\mathbf{x}\right) \tilde \varTheta^{u}_g\left(\mathbf{x}\right) \varTheta^{d\dagger}_g\left(\mathbf{x}\right)
 \left|A\left(\mathbf{x}\right)\right\rangle
 \label{eqsym1}
\end{equation}
(see figure \ref{fig1}d).
This equation may be interpreted as a Gauss law for a lattice gauge theory whose gauge group is $G$ \cite{Zohar2015}, \emph{with virtual ``electric'' fluxes on the links} which are not physically measurable.

The global physical symmetry is guaranteed since
\begin{multline}
\Theta_g\left|\psi\right\rangle  = \left\langle\left\{HV\right\}\right|
\underset{\mathbf{x}}{\bigotimes} \Theta_g^p\left(\mathbf{x}\right)\left|A\left(\mathbf{x}\right)\right\rangle=\\
= \left\langle\left\{HV\right\}\right| \underset{\mathbf{x}}{\bigotimes} \varTheta^{l \dagger}_g\left(\mathbf{x}\right) \tilde \varTheta^{r }_g\left(\mathbf{x}\right)
 \tilde \varTheta^{u }_g\left(\mathbf{x}\right) \varTheta^{d\dagger}_g\left(\mathbf{x}\right) \left|A\left(\mathbf{x}\right)\right\rangle = \left|\psi\right\rangle,
\end{multline}
as the state $\left|\left\{HV\right\}\right\rangle$ is invariant under $G$ (see Eq. (\ref{eqinv})).

The condition (\ref{eqsym1}) is satisfied if one chooses, for example,
\begin{equation}
A^{j_p m_p}_{j_l m_l;j_r m_r;j_u m_u;j_d m_d} = \underset{j_1 j_2}{\sum}\alpha^{j_p j_1 j_2}_{j_l j_r j_u j_d}
\left\langle j_l m_l j_d m_d | j_1 m_1 \right\rangle
 \left\langle j_1 m_1 j_p m_p | j_2 m_2 \right\rangle \left\langle j_2 m_2 | j_r m_r j_u m_u \right\rangle \label{CGtensor}
\end{equation}
where $\alpha^{j_p j_1 j_2}_{j_l j_r j_u j_d}$ are free (variational) parameters and the coefficients $\bracket{j_a m_a j_b m_b}{j_c m_c}$ are the Clebsch-Gordan coefficients of the group. This expression generalizes the construction of the vertex presented in \cite{Tagliacozzo2014} to include also the matter states. The proof  of the gauge invariance of $A$ can be found in the Appendix and a graphical representation of the tensor is depicted in Fig. \ref{fig:altA}. We emphasize that the Clebsch-Gordan coefficients must account also for the presence of possible degeneracies $d_j$ of the representations. As previously discussed $j$ and $m$ indices are potentially sets of multiple indices.

If one wishes to increase the bond dimension, further copies of the virtual irreducible representations may be used (see Sec. \ref{sec:Apr}). On the other hand, the dimensions of the
required Hilbert spaces may be reduced if one chooses only a few representations, as long as all the states of every chosen representation are considered, and the representations chosen are connected by nonvanishing Clebsch-Gordan coefficients \cite{Zohar2015}.

As an example, consider the simple Abelian case where $G=U(1)$. The irreducible representations are all one-dimensional, therefore the representation index $j$ is irrelevant, and the Clebsch-Gordan coefficients are simply
$\left\langle m_1 m_2 | m \right\rangle = \delta_{m_1+m_2,m}$. Thus one obtains that $A^{m_p}_{m_l m_r m_u m_d} = \alpha^{m_p}_{m_l m_r m_u m_d} \delta_{m_r+m_u-m_l-m_d,m_p}$ from which the Abelian Gauss law with a physical charge $m_p$ on the vertex and virtual fluxes $m_{l,r,u,d}$ on the links around it is easily recognizable.

\section{Gauging the symmetry}
The next step is to \emph{gauge the symmetry}, or to make it local, i.e. to change our PEPS such that it will be locally invariant under $G$: we shall define a local unitary
 transformation $\Theta_g\left(\mathbf{x}\right)$ acting on the physical Hilbert space, such that for every vertex $\mathbf{x}$,
 \begin{equation}
 \Theta_g\left(\mathbf{x}\right) \left|\psi\right\rangle = \left|\psi\right\rangle.
 \label{Lsym}
 \end{equation}
 This is a local symmetry, in the sense that the above equation holds for any vertex $\mathbf{x}$ independently of the other vertices: one could choose different group elements to act with on different vertices ($g \rightarrow g\left(\mathbf{x}\right)$).

 For that, we will need to introduce new physical degrees of freedom, residing on the links of the lattice. Since we have already interpreted the virtual states as eigenstates of some virtual electric flux (or field), and the PEPS symmetry (\ref{eqsym1}) as a Gauss law, it would be natural now to ``physicalize'' the virtual states, i.e. to add physical degrees of freedom on each link whose transformations rules are identical to the virtual ones, and will satisfy a physical Gauss law.  For that, we introduce a new type of tensors, \emph{Gauge Tensors}, residing on the links.

\subsection{Gauge tensors}
\begin{figure}[th]
  \centering
  \includegraphics[width=\textwidth]{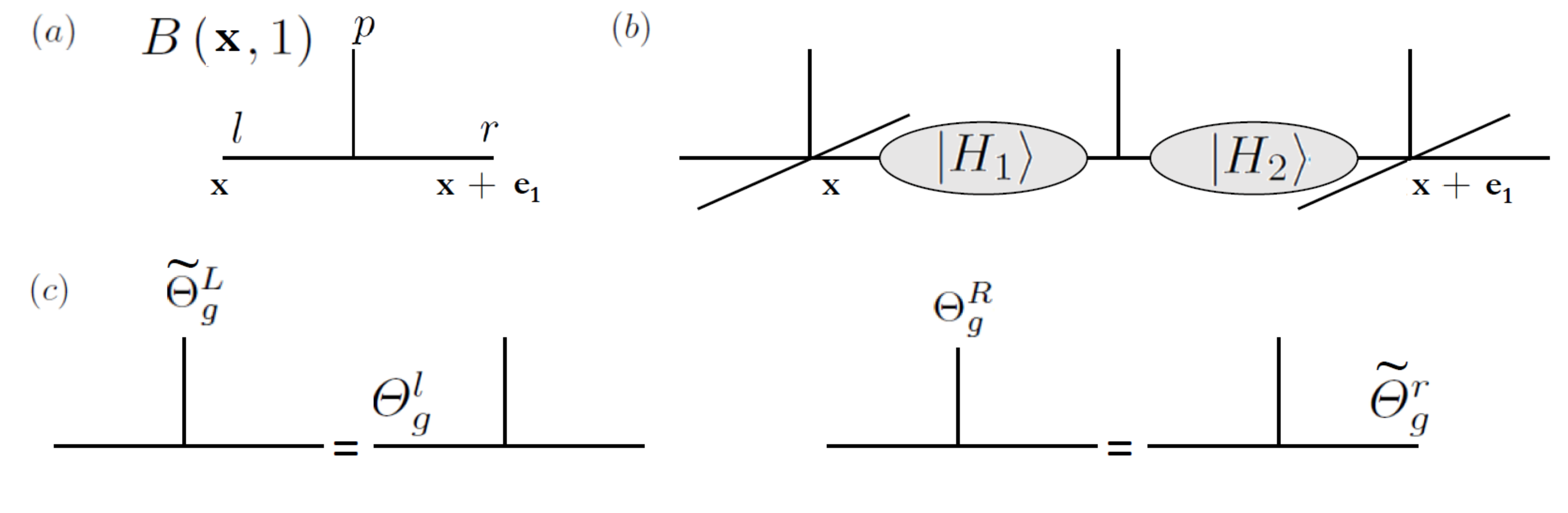}\\
  \caption{The $B$ tensor and its symmetries. Only the horizontal case is drawn; the vertical link tensors are similar.
   (a) The link tensor, with the physical state $p$ and the virtual states $l,r$, as defined
  in (\ref{eqBten}). (b) Projection to entangled states on the bonds. (c) The symmetries of the fiducial state (\ref{eqsym2}). }\label{fig2}
\end{figure}

We denote each link by two indices: $\mathbf{x}$, the lattice vertex from which it emanates (rightwards or upwards), and $k$, its direction. On each link we define a fiducial state based on a new kind of tensors, $B^{p}_{ab}$,
\begin{equation}
\left|B\left(\mathbf{x},k\right)\right\rangle = B^{p}_{ab}\left|p\right\rangle\left|ab\right\rangle
\label{eqBten}
\end{equation}
where $a=l$, $b=r$ for $k=1$ (horizontal links) and $a=d$, $b=u$ for $k=2$ (vertical links) - see figure \ref{fig2}a. Here $\ket{p}$ is a physical state that is associated to a gauge field in the lattice gauge theory framework.

One needs to modify the  entangled bond states to connect the $B$ and the $A$ tensors. For that, we define two projectors on each link, connecting each $B$ tensor to the two neighboring $A$ tensors (see Figure \ref{fig2}b). For example, instead of $\left|H\right\rangle$ ($\left|V\right\rangle$) we will now project to two states,
$\left|H_1\right\rangle$ ($\left|V_1\right\rangle$) connecting  $A$ from its right (upper) edge  with $B$, and $\left|H_2\right\rangle$ ($\left|V_2\right\rangle$), connecting $A$ from its left (lower) edge  to $B$. These states define a new, generalized $\left|\left\{HV\right\}\right\rangle$, under the assumption that the virtual Hilbert spaces of $B$ match those of $A$.

In order to keep the new bond states invariant under group transformations, all we have to do is to demand that the left/down virtual states of the links will transform with a \emph{right transformation rule}, while the right/up transform with the \emph{left} ones. Then, a result similar to (\ref{eqinv}) is immediate for both the horizontal and vertical links.
The physical states for the gauge field on the links will take the form $\left|p\right\rangle = \left|j_p m_p n_p\right\rangle$. These are \emph{representation states} \cite{Zohar2015}, following
both a left ($m_p$) and a right ($n_p$) transformation rules - the physical transformations are
\begin{equation}
\tilde \Theta^{L/D}_g \left|j_p m_p n_p\right\rangle = D^{j}_{m_p m'_p}\left(g\right)\left|j_p m'_p n_p\right\rangle
\end{equation}
and
\begin{equation}
\Theta^{R/U}_g \left|j_p m_p n_p\right\rangle = D^{j}_{n'_p n_p}\left(g\right)\left|j_p m_p n'_p\right\rangle.
\end{equation}
(as described in section \ref{sec:math}).
We define the fiducial product state
\begin{equation}
\left|\Phi\left(A,B\right)\right\rangle =
\underset{\mathbf{x}}{\bigotimes}\left|A\left(\mathbf{x}\right)\right\rangle
\underset{\mathbf{x},k}{\bigotimes}\left|B\left(\mathbf{x},k\right)\right\rangle \equiv \left|\Phi\left(A\right)\right\rangle\left|\Phi\left(B\right)\right\rangle
\end{equation}
such that the physical state that includes both vertex and link degrees of freedom reads:
\begin{equation}
\left|\psi\right\rangle = \left\langle\left\{HV\right\} | \Phi\left(A,B\right)\right\rangle.
\end{equation}

Following the ``physicalization'' of the Gauss law (\ref{eqsym1}), we wish the local physical transformation around the vertex $\mathbf{x}$ to be:
\begin{equation}
 \Theta_g\left(\mathbf{x}\right) =  \Theta_g^p\left(\mathbf{x}\right)
\tilde \Theta^{L\dagger}_g\left(\mathbf{x},1\right)
\tilde \Theta^{D\dagger}_g\left(\mathbf{x},2\right)
\Theta^{R}_g\left(\mathbf{x-\hat e}_1,1\right)
\Theta^{U}_g\left(\mathbf{x-\hat e}_2,2\right).
\label{GaussOp}
\end{equation}

We have defined all the ingredients required for eq. (\ref{Lsym}), and thus we can finally turn to its fulfillment. For that, what we need is to impose, on top of the symmetry requirements on $A$ (\ref{eqsym1}), two more similar symmetry requirements on $B$:
\begin{equation}
\begin{aligned}
& \tilde \Theta^{L/D}_g\left(\mathbf{x},k\right)\left|B\left(\mathbf{x},k\right)\right\rangle =
\varTheta^{l/d}_g\left(\mathbf{x},k\right)\left|B\left(\mathbf{x},k\right)\right\rangle \\
& \Theta^{R/U}_g\left(\mathbf{x},k\right)\left|B\left(\mathbf{x},k\right)\right\rangle =
 \tilde \varTheta^{r/u}_g\left(\mathbf{x},k\right)\left|B\left(\mathbf{x},k\right)\right\rangle
\label{eqsym2}
\end{aligned}
\end{equation}
(see figure \ref{fig2}c).

If the $B$ tensors satisfy these properties, the local gauge symmetry is guaranteed. Let us verify it:
\begin{equation}
\begin{aligned}
 \Theta_g\left(\mathbf{x}\right) \left|\psi\right\rangle =&
\left\langle \left\{HV\right\}\right|\Theta^p_g\left(\mathbf{x}\right) \left|\Phi\left(A\right)\right\rangle
\tilde \Theta^{L\dagger}_g\left(\mathbf{x},1\right)
\tilde \Theta^{D\dagger}_g\left(\mathbf{x},2\right) \times \\ & \hspace{1cm}
 \Theta^{R}_g\left(\mathbf{x-\hat e}_1,1\right)
\Theta^{U}_g\left(\mathbf{x-\hat e}_2,2\right) \left|\Phi\left(B\right)\right\rangle = \\
 &\left\langle \left\{HV\right\}\right|
\tilde \varTheta^{r}_g\left(\mathbf{x}\right) \varTheta^{l \dagger}_g\left(\mathbf{x},1\right)
\tilde \varTheta^{u}_g\left(\mathbf{x}\right) \varTheta^{d \dagger}_g\left(\mathbf{x},2\right) \times \\& \hspace{1cm}
 \varTheta^{l\dagger}_g\left(\mathbf{x}\right) \tilde \varTheta^{r}_g\left(\mathbf{x-\hat e}_1,1\right)
\varTheta^{d\dagger}_g\left(\mathbf{x}\right) \tilde \varTheta^{u}_g\left(\mathbf{x-\hat e}_2,2\right) \left|\Phi\left(A,B\right)\right\rangle =
\left|\psi\right\rangle
\end{aligned}
\end{equation}
where the second equality follows from the symmetry requirements (\ref{eqsym1},\ref{eqsym2}), and the third one - from the invariance of the maximally
entangled bond states under $G$.

It is straightforward to see that a general $B$ tensor which satisfies these requirements is
\begin{equation} \label{eqBten2}
B^{j_p m_p n_p}_{j_{l/d} m_{l/d};j_{r/u} m_{r/u}}=\beta^{j_p} \delta_{j_p j_{l/d}} \delta_{j_p j_{r/u}} \delta_{m_p m_{l/d}} \delta_{n_p m_{r/u}}.
\end{equation}

Just like the $A$ tensors, one may also truncate $B$ and include only some representations, as long as the truncation scheme discussed in \cite{Zohar2015} is fulfilled.

Note that the gauging procedure described above is quite general, and depends only on using, for the $B$ tensors, the same virtual subspace as the one of the $A$ tensors, and the fulfillment of conditions (\ref{eqsym2}); this is completely independent of the choice of the $A$ tensors, given that they satisfy the symmetry requirement (\ref{eqsym1}), and the bond states satisfy the respectively required invariance properties.

In the simple case of $G=U(1)$, one does not need to consider separate $m$ and $n$ values, as $m=n$. Then, simply, on top of the $U(1)$ Gauss law of $A$,
here one obtains that $B^{m_p}_{m_{l/d} m_{r/u}} = \beta^{m_p} \delta_{m_p,m_{l/d}} \delta_{m_p,m_{u/r}}$.

\subsection{Unification of Tensors}
\begin{figure}
  \centering
  \includegraphics[width=\textwidth]{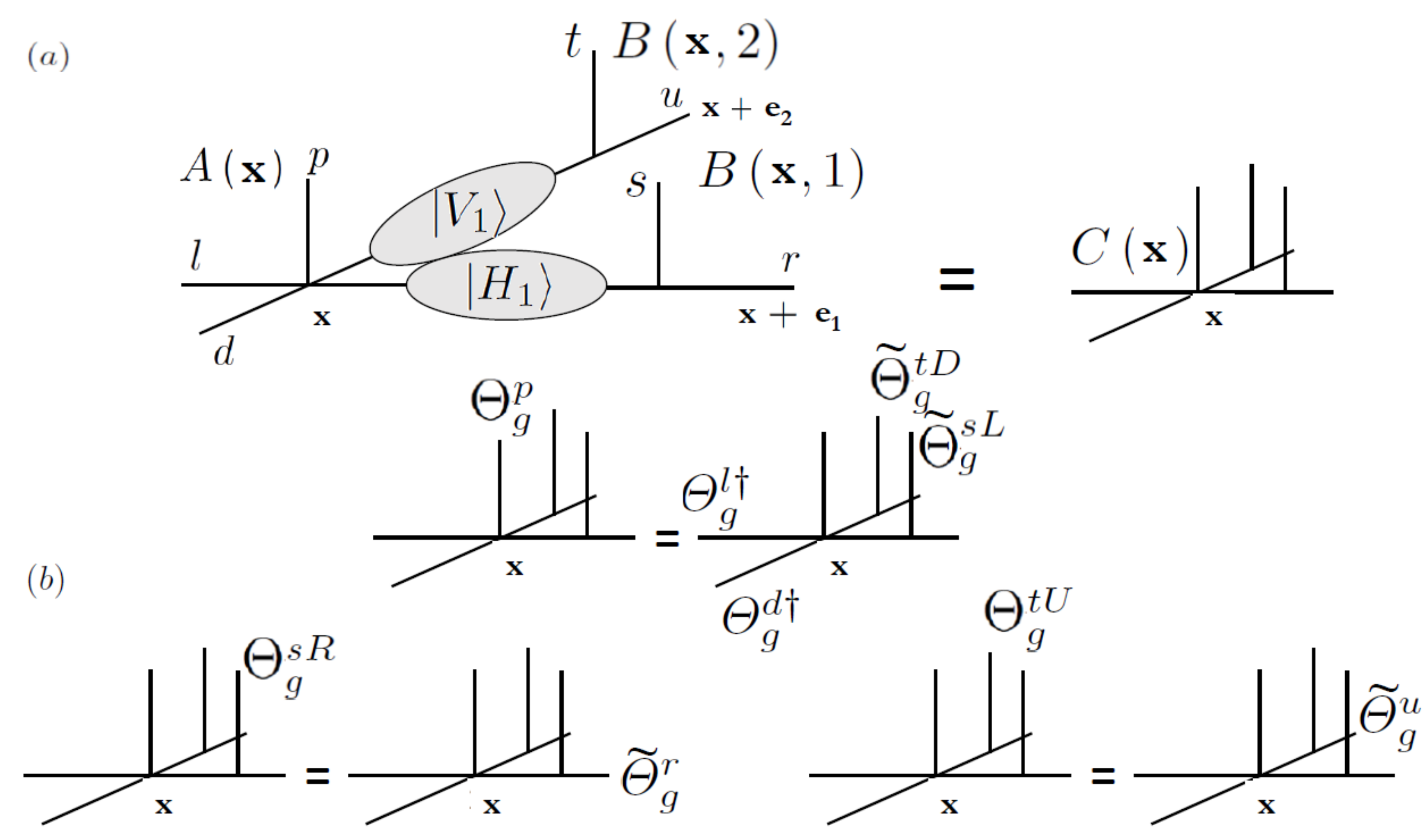}\\
  \caption{The $C$ tensor and its symmetries. (a) The definition of $C$, based on the previously defined $A$ and $B$ (\ref{eqCdef}).
  (b) The symmetry properties of $C$ (\ref{eqsym3}).}\label{fig3}
\end{figure}

The $B$ tensors are very important for a constructive derivation of a locally gauge invariant PEPS, but actually one does not need two types of tensors
to get such a state. In fact, the $A$,$B$ tensors may be united into a single tensor $C$, residing on the vertices, consisting of four virtual states
$l,r,u,d$ and three physical states $p,s,t$, where $p$ is the physical state of $A$, and $s,t$ (standing for ``side'' and ``top'') are the physical states of the $B$ tensors from the right and the top neighboring links of the vertex. Mathematically,
\begin{equation}
\left|C\left(\mathbf{x}\right)\right\rangle = \left(\left\langle \left\{H_1\left(\mathbf{x};\mathbf{x},1\right)\right\}\right|
\otimes \left\langle \left\{V_1\left(\mathbf{x};\mathbf{x},2\right)\right\}\right| \right)
 \left(\left|A\left(\mathbf{x}\right)\right\rangle \otimes \left|B\left(\mathbf{x},1\right)\right\rangle \otimes \left|B\left(\mathbf{x},2\right)\right\rangle \right)
\label{eqCdef}
\end{equation}
(see figure \ref{fig3}a), or, simply,
\begin{equation}
\left|C\left(\mathbf{x}\right)\right\rangle = \underset{\left\{j\right\}}{\sum} C^{j_p m_p;j_s m_s n_s;j_t m_t n_t}_{j_l m_l;j_r m_r;j_u m_u;j_d m_d}
\left|j_p m_p;j_s m_s n_s;j_t m_t n_t\right\rangle \left|j_l m_l;j_r m_r;j_u m_u;j_d m_d\right\rangle
\end{equation}
where
\begin{multline}
 C^{j_p m_p;j_s m_s n_s;j_t m_t n_t}_{j_l m_l;j_r m_r;j_u m_u;j_d m_d} \equiv A^{j_p m_p}_{j_l m_l;j_s m_s;j_t m_t;j_d m_d}
 B^{j_s m_s n_s}_{j_s m_s;j_r m_r}B^{j_t m_t n_t}_{j_t m_t;j_u m_u} = \\
 = A^{j_p m_p}_{j_l m_l;j_s m_s;j_t m_t;j_d m_d} \beta^{j_s} \delta_{j_s,j_r} \delta_{n_s,m_r} \beta^{j_t} \delta_{j_t,j_u} \delta_{n_t,m_u}
\end{multline}
here, no sum is intended for repeated indices, and the last equality is obtained by Eq. \eqref{eqBten2} and links the physical indices $j_s,n_s,j_t,n_t$ with the virtual indices $j_r,m_r,j_u,m_u$ respectively.

The only remaining bond states are the ones previously denoted $\left|H_2\right\rangle,\left|V_2\right\rangle$.

The symmetry relations (\ref{eqsym1},\ref{eqsym2}) are now simply mapped to
\begin{equation}
\begin{aligned}
&\Theta^{p}_g\left(\mathbf{x}\right) \left|C\left(\mathbf{n}\right)\right\rangle =
\varTheta^{l \dagger}_g\left(\mathbf{x}\right) \tilde \varTheta^{s,L}_g\left(\mathbf{x}\right) \tilde \varTheta^{t,D}_g\left(\mathbf{x}\right) \varTheta^{d\dagger}_g\left(\mathbf{x}\right)
 \left|C\left(\mathbf{x}\right)\right\rangle\,, \\
& \tilde \Theta^{sR/tU}_g\left(\mathbf{x},k\right)\left|C\left(\mathbf{x},k\right)\right\rangle =
\varTheta^{r/u}_g\left(\mathbf{x},k\right)\left|C\left(\mathbf{x},k\right)\right\rangle
\label{eqsym3}
 \end{aligned}
 \end{equation}
 (see figure \ref{fig3}b).

\subsection{Physical description of the states}
Let us now elaborate a little more on the physical meaning of these states.
Our PEPS are simply states of a lattice gauge theory, whose gauge group is $G$. The physical degrees of freedom on the vertices are simply the matter
fields, with charges $\left|j_p m_p\right\rangle$. The physical states on the links, $\left|j m n\right\rangle$, are simply eigenstates of the (generally non-Abelian left and right) electric field \cite{KogutSusskind,Zohar2015}. The Gauss law, as we have already argued, simply corresponds to Eq. (\ref{Lsym}).

In the case in which $G$ is a compact Lie group, as explained in \ref{sec:math} the transformation operators $\Theta_g$ may be written in terms of parameters and generators. For example, $\Theta^{p} = e^{i \mathbf{q}_g \cdot \mathbf{Q}}$, $\tilde \varTheta^{L} = e^{i \mathbf{q}_g \cdot \mathbf{L}}$,
$\varTheta^{R} = e^{i \mathbf{q}_g \cdot \mathbf{R}}$, where $\mathbf{q}_g$ is the vector of group parameters for $g$, and $\mathbf{Q,L,R}$ are the generators of transformations for the matter (charge) and the gauge field (left and right electric fields), satisfying the group algebra
\begin{equation}
\begin{aligned}
& \left[Q_a,Q_b\right]=if_{abc}Q_c \\ & \left[L_a,L_b\right]=-if_{abc}L_c \\  & \left[R_a,R_b\right]=if_{abc}R_c \\ & \left[L_a,R_b\right]=0
\end{aligned}
\end{equation}
($f_{abc}$ are the structure constants of $G$).

We can also find the generators $G\left(\mathbf{x}\right)$ of the gauge transformations $\Theta_g\left(\mathbf{x}\right)$; it is straightforward to deduce from equation (\ref{Lsym}) that the Gauss law for the generators is simply
\begin{equation}
\mathbf{G}\left(\mathbf{x}\right)\left|\psi\right\rangle \equiv \big(
\mathbf{Q}\left(\mathbf{x}\right) - \mathbf{L}\left(\mathbf{x},1\right) - \mathbf{L}\left(\mathbf{x},2\right)
+ \mathbf{R}\left(\mathbf{x-\hat e}_1,1\right) + \mathbf{R}\left(\mathbf{x-\hat e}_2,2\right)
\big) \left|\psi\right\rangle = 0
\end{equation}
which is recognizable as the discrete divergence of the lattice Gauss law.
Furthermore, since eq. (\ref{Lsym}) does not allow the introduction of any additional phase, we obtain that $\left|\psi\right\rangle$ is in the kernel of $G\left(\mathbf{x}\right)$ and thus there are no static charges.

\section{Properties of the vertex tensors} \label{sec:Apr}
As shown in \ref{app:A}, the gauge invariance of the physical states is derived from the particular form chosen for the tensor $A$ in Eq. \eqref{CGtensor}. For the sake of simplicity, so far we have considered the parameters $\alpha^{j_pj_1j_2}_{j_lj_rj_uj_d}$ as single parameters. This provides a minimal requirement to obtain the gauge invariance of the states, eventually considering the truncation to suitable representations.

The structure presented in \eqref{CGtensor}, though, can be generalized to accomodate a larger number of variational parameters that are necessary for the purpose of using our PEPS construction as a variational family to numerically investigate the properties of systems whose dynamics is dictated by a gauge-invariant lattice Hamiltonian.

With this aim, the parameters $\alpha^{j_pj_1j_2}_{j_lj_rj_uj_d}$ can be promoted to matrices acting on a larger space of virtual states. To this purpose we must extend the basis to describe the virtual states along each link: each virtual state of the form $\ket{jm}$ can be substituted with a new Hilbert ``degeneracy space'' generated by $\left\lbrace \ket{jm,i}\right\rbrace $ with $i=1,\ldots,\tilde{d}_{j}$ an index which distinguish $\tilde{d}_{j}$ orthogonal states (the virtual states $\ket{jm,i}$ must not be confused with the physical states $\ket{jmn}$ which represents the gauge fields). In this way we are increasing the bond dimension of the PEPS construction. We emphasize that the degeneracy $\tilde{d}_{j}$ is only a technical feature of the construction of the tensor network and it is not related with the possible physical degeneracies $d_j$ of the representations of the group discussed in Sec. \ref{sec:math}. In particular, each of the states $\ket{jm,i}$ obeys the same transformation rules \eqref{grouptransformation}, but now, each element of $D^j_{mm'}(g)$ must be extended to an arbitrary unitary operator $(D^j_{mm'}(g))_{ii'}$ that maps the basis $\left\lbrace \ket{jm,i}\right\rbrace $ into the basis $\left\lbrace \ket{jm',i'}\right\rbrace$. This approach constitutes indeed in adding a given $\tilde{d}_j$-degeneracy at the virtual level for each representation of the group. In this way each representation is considered $\tilde{d}_j$ times on the virtual links and the operators $\varTheta_g$ must be generalized accordingly. The entangled states $H$ and $V$ must account for the new degeneracies as well.

Finally, the tensor $B$ must be suitably extended as well. In particular, the parameter $\beta^{j_p}$ in Eq. \eqref{eqBten2} is now lifted to a $\tilde{d}_{j_p} \times \tilde{d}_{j_p}$ tensor of variational parameters. The tensor $C$ must be modified accordingly and is still described by the construction \eqref{eqCdef}.

An analogous approach to the construction of variational tensor network states with degeneracy spaces has been discussed in detail in \cite{Tagliacozzo2014} for the case of pure gauge theories.

Beside this additional degeneracy of the virtual states, there is also an intrinsic degeneracy of the tensor $A$ that can be exploited: it is the one given by the ``inner indices'' $j_1$ and $j_2$ in $\alpha^{j_pj_1j_2}_{j_lj_rj_uj_d}$. $j_1$ and $j_2$ label, from the perspective of the fusion algebra of the irreducible representations of $G$, are independent fusion channels which can be separately addressed.

\begin{figure}
 \includegraphics[width=\textwidth]{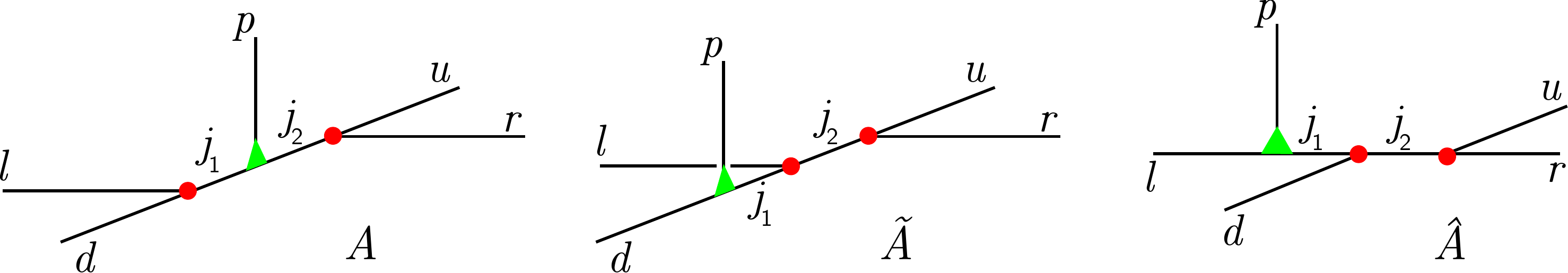}
 \caption{The possible formulations for the vertex tensor, $A, \tilde{A}$ and $\hat{A}$, are depicted (see Eqs. \eqref{CGtensor} and \eqref{altA}). They differ by the ordering of the fusion of the representations of the external virtual modes $l,d,r,u$ with the physical mode $p$. Therefore the inner indices $j_1$ and $j_2$ assume a different meaning in the three cases which amount to three different basis for the state defined by the fusion of the five external fermionic states. The red dots in the figure represent Clebsch-Gordan coefficients relating virtual states with the inner indices $j_1$ and $j_2$, whereas the green triangles depict Clebsh-Gordan coefficients involving also physical states. The ordering of the fusions is meant to be from left to right.} \label{fig:altA}
\end{figure}

Let us analyze more closely the role of these indices. $\left|j_1 m_1\right\rangle$ is an intermediate state of the coupling of three group states (such as the addition of three angular momenta in the $SU(2)$ case, for example), which can be performed in three different ways (depending on the two states which one chooses to add first, before adding to the third one); therefore it is possible to give different definitions of the tensor $A$ by adopting the two other options. In particular, by changing the order of the fusion of the first three representations, $j_l,j_d$ and $j_p$, we can use two alternative formulations for $A^{j_p m_p}_{j_l m_l;j_r m_r;j_u m_u;j_d m_d}$ (see Fig. \ref{fig:altA} for a pictorial representation):

\begin{align} \label{altA}
 \tilde{A}^{j_p m_p}_{j_l m_l;j_r m_r;j_u m_u;j_d m_d} &= \tilde{\alpha}^{j_p j_1 j_2}_{j_l j_r j_u j_d} \bracket{j_d m_d j_p m_p}{j_1 m_1}\bracket{j_l m_l j_1 m_1}{j_2 m_2} \bracket{j_2 m_2}{j_r m_r j_u m_u}\, \\
 \hat{A}^{j_p m_p}_{j_l m_l;j_r m_r;j_u m_u;j_d m_d} &= \hat{\alpha}^{j_p j_1 j_2}_{j_l j_r j_u j_d} \bracket{j_l m_l j_p m_p}{j_1 m_1}\bracket{j_1 m_1 j_d m_d}{j_2 m_2} \bracket{j_2 m_2}{j_r m_r j_u m_u}\,
\end{align}

Besides the external representations, the coefficients $\alpha,\tilde{\alpha}$ and $\hat{\alpha}$ depend on the two internal fusion representations $j_1$ and $j_2$, which assume a different physical meaning in the three cases. Therefore, independently of the chosen form for the tensor $A$, the coefficients $\alpha,\tilde{\alpha}$ and $\hat{\alpha}$ can be freely chosen as a function of $j_1$ and $j_2$ without breaking the gauge invariance. The three choices, however, are not independent. The properties of the group fix indeed the relations between the different orderings that we adopted to represent the fusion of the first three external group states $\ket{j_l m_l} \otimes \ket{j_d m_d} \otimes \ket{j_p m_p}$. The three choices correspond to three different bases representing the same Hilbert space. In particular one can introduce the so-called $6j-$symbols, $\mathcal{F}^{j_l j_d j_p}_{j_2 j_1 j_1^\prime}$ which constitute suitable mappings among these bases. For example we have (see, for instance, \cite{Rose1995}, for the $SU(2)$ case):
\begin{equation} \label{eqF}
ì\bracket{j_l m_l j_d m_d}{j_1 m_1} \bracket{j_1 m_1 j_p m_p}{j_2 m_2} = \sum_{j_1^\primeì} \mathcal{F}^{j_l j_d j_p}_{j_2 j_1 j_1^\prime} \bracket{j_d m_d j_p m_p}{j_1^\prime m_1^\prime}\bracket{j_l m_l j_1^\prime m_1^\prime}{j_2 m_2}
\end{equation}
where, as usual, the summation over repeated $m$ indices is implied. Therefore:
\begin{equation}
 \alpha^{j_p j_1 j_2}_{j_l j_r j_u j_d} = \sum_{j_1^\prime} \mathcal{F}^{j_l j_d j_p}_{j_2 j_1 j_1^\prime} \tilde{\alpha}^{j_p j_1^\prime j_2}_{j_l j_r j_u j_d}
\end{equation}
To express $\alpha$ in terms of the coefficients $\hat{\alpha}$, it is also necessary to introduce a matrix $\mathcal{B}$ which describes the exchange of two representations fusing in a Clebsch-Gordan coefficient. For the case of SU(2), such matrix does not depend on the group indeces $m$ and we have:
\begin{equation}
 \bracket{j_a m_a j_b m_b}{j_c m_c} = \mathcal{B}^{j_a j_b}_{j_c } \bracket{j_b m_b j_a m_a}{j_c m_c}\,,
\end{equation}
with $\mathcal{B}^{j_a j_b}_{j_c}=(-1)^{j_a+j_b-j_c}$. In this case, by combining $\mathcal{F}$ and $\mathcal{B}$ transformations we get:
\begin{equation} \label{finaltransform}
 \alpha^{j_p j_1 j_2}_{j_l j_r j_u j_d} = \sum_{j_1^\prime} \mathcal{F}^{j_l j_d j_p}_{j_2 j_1 j_1^\prime} \tilde{\alpha}^{j_p j_1^\prime j_2}_{j_l j_r j_u j_d}=\sum_{j_1^\prime,j_1^{\prime\prime}} \mathcal{F}^{j_l j_d j_p}_{j_2 j_1 j_1^\prime} \mathcal{B}_{j_d j_p}^{j_1^\prime}(\mathcal{F}^{-1})^{j_l j_p j_d}_{j_2j_1^\prime j_1^{\prime \prime}}\hat{\alpha}^{j_p j_1^{\prime\prime} j_2}_{j_l j_r j_u j_d}\,.
\end{equation}
Similar relations hold for the more general case in which $\mathcal{B}$ depends on the $m$ indices as well.

The previous equations are the transformations related to the first representation fusion $\ket{j_l m_l} \otimes \ket{j_d m_d} \otimes \ket{j_p m_p}$ but, of course, similar arguments can be extended also to involve the outgoing representations $\ket{j_r m_r} \otimes \ket{j_u m_u}$ and, by mixing ingoing and outgoing representations, one can obtain further equivalent parameterizations of $A$ involving also transformations of the other inner index $j_2$. In this case, however, due to the different gauge transformation rules of the right and up states, additional care must be devoted in writing the $\mathcal{F}$ matrices, which must be rewritten to account for the conjugate of the Clebsch-Gordan coefficients.

Let us observe here that, similarly to the virtual bond modes, we can impose a truncation in the representations of the inner indices $j_1$ and $j_2$ as well. This corresponds to a mapping from a gauge group to the corresponding quantum groups \cite{Gomez1996} and it requires a suitable modification the Clebsch-Gordan coefficients which must be transformed in quantum Clebsch-Gordan coefficients. Furthermore, if we do so, the $\mathcal{F}$ matrices of the theory must be corrected accordingly, in such a way that, by changing the order of the fusions appearing in the definition of the tensor $A$, all the possible inner fusion outcomes fulfill the same truncation.

\begin{figure}[hbt]
 \includegraphics[width=\textwidth]{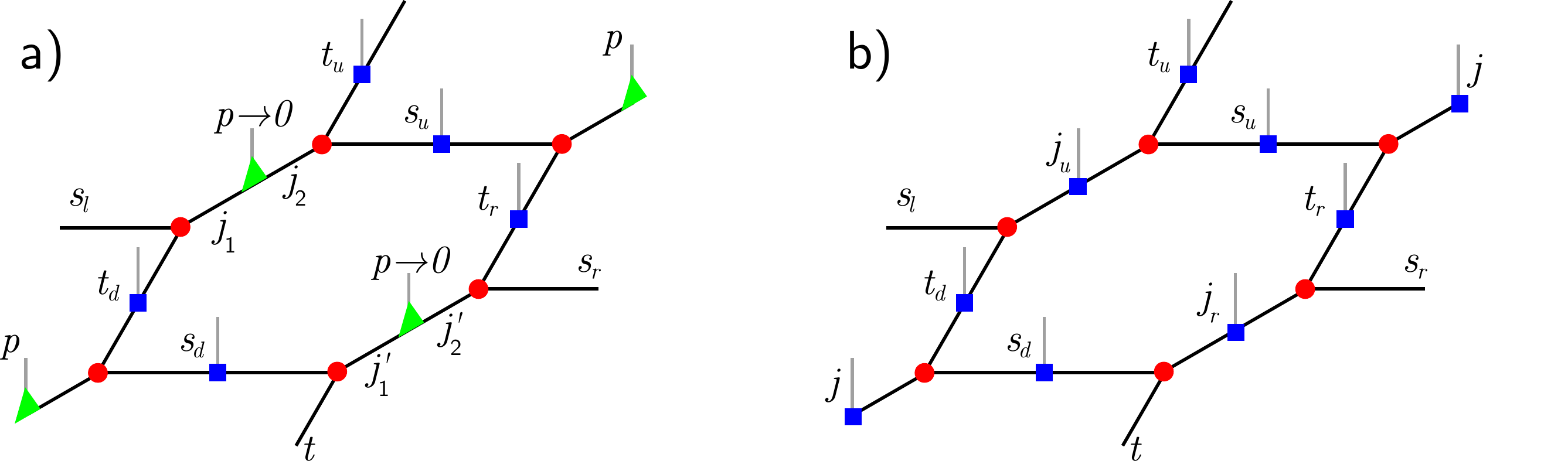}
 \caption{Transformation from a gauge invariant state with physical matter and gauge bosons on a square lattice into a topological state with physical bosons only on a honeycomb lattice. a) A plaquette of the PEPS for the lattice gauge theory on a square lattice is represented. The gauge bosons $s_u,t_r,s_d,t_d$ live on the four edges of a square plaquette. The $j$'s represent only the inner irreducible representations needed to express the tensor $A$ in Eq. \eqref{CGtensor}. When the physical fermionic states $p$ are set to the vacuum state, then $j_1=j_2$ and $j_1'=j_2'$. The red dots represent Clebsch-Gordan coefficients involving virtual states only, whereas the green triangles are Clebsch-Gordan coefficients referring to the physical matter states which define the vertices of the lattice plaquette. b) To obtain a topological PEPS reminiscent of a string-net model on the honeycomb lattice, we substitute the physical fermionic degrees of freedom $p$ with new gauge bosons $j$. In particular we must introduce additional $B$ tensors (see Eq. \eqref{eqBten2}) for the new ``physicalized'' degrees of freedom $j$ whose representations are related to the previous inner representations by $j_u=j_1=j_2$ and $j_r=j_1'=j_2'$. The lattice is now a honeycomb lattice and the plaquette is defined by the six physical states $j_u,s_u,t_r,j_r,s_d,t_d$. In b) the red dots represent quantum Clebsch-Gordan coefficients. In both panels the blue squares depict $B$ tensors and the projectors are not illustrated for the sake of simplicity.} \label{fig:honeycomb}
\end{figure}

For the specific case of $SU(2)$ this procedure is consistent with strengthening the $SU(2)$ Gauss law in such a way that all the fusions of three representations fulfill the rules of a truncated $SU(2)_k$ algebra, where $k$ specifies the truncation and each representation is limited to $j=0,\ldots,k/2$ (see \cite{Gomez1996} for details on quantum groups and their truncated representations). This is the first step required for obtaining a tensor network state with topological order in the spirit of string-net models \cite{Gu2009,Aguado2009}. In particular, after the introduction of the link physical states $s$ and $t$, we can set all the physical matter states to $j_p=0$, in such a way that we obtain the condition $j_1=j_2$ in the inner indices of the tensors $A$. This new construction gives a new honeycomb lattice where each edge is associated either to a representation carried by a physical link state, or to a representation carried by an inner $j_1=j_2$ edge. By promoting the latter to physical states as well, through the introduction of suitable $B$ tensors, we find a state that satisfies all the fusion requirements of a string-net model in a honeycomb lattice (see Fig. \ref{fig:honeycomb}).

\section{Gauge-invariant fermionic PEPS} \label{sec:fermions}
One may also use fermionic PEPS (fPEPS) \cite{Kraus2010} for the creation of locally gauge invariant states. These are particularly useful in the cases where one wishes to describe the dynamic charges in terms of dynamical fermions, as in the standard model of particle physics.

The main difference is in the definition of the fiducial state $\left|A\right\rangle$; since we are dealing with fermions here, one has to consider the  fermionic algebra and statistics, which arise in this context. For example, the tensor product structure of Hilbert spaces used along the PEPS construction in the previous sections has to be reworked, since tensor products are not defined for fermionic Hilbert spaces (one has to perform an appropriate antisymmetrization).

The main strategy is then to exploit a second-quantized approach that allows us to describe the fiducial states and projectors in terms of fermionic operators acting within a suitable Fock space \cite{Kraus2010}. In this way the anticommutation relations between operators are naturally satisfied.
Keeping in mind this difference, we will follow an approach similar to the previous sections  to build the gauge-invariant states with fermionic matter as well: we will start from a global symmetry, in a purely fermionic systems and then we will localize it, by adding suitable physical bosonic degrees of freedom in the links, which will play the role of the gauge fields.

\subsection{Fermionic PEPS with a global symmetry}
First, we consider the fermionic fiducial states $\left|A\right\rangle$ and the projectors connecting them to a physical state. We denote the fermionic vacuum by $\left|\Omega\right\rangle$, and the physical and virtual vacua by $\left|\Omega_p\right\rangle$, $\left|\Omega_v\right\rangle$ respectively. We will sometimes use these notations for either local and global vacua, as will be clear from the context;
we emphasize, however, that the vacuum must be intended as a Fock vacuum rather than a product state. Thus, for example, the vertex fiducial state of physical and virtual fermions, before the projections, is
\begin{equation}
\left|\Phi\left(A\right)\right\rangle = \underset{\mathbf{x}}{\prod}\mathcal{A}\left(\mathbf{x}\right)\left|\Omega\right\rangle
\end{equation}
where $\mathcal{A}$ is a second quantized operator composed of fermionic creation operators which acts on the global Fock vacuum state $\ket{\Omega}$.

We want the operators $\mathcal{A}\left(\mathbf{x}\right)$ to commute with one another for different vertices $\mathbf{x}$, and thus they should all have an even fermionic parity, considering  physical and virtual fermions together; this  requirement will be satisfied by all the following examples.

More generally, the fermionic parity plays an important role in the construction of the fiducial states of fermionic PEPS \cite{Kraus2010}. This becomes evident when we want to combine fermionic matter, local gauge invariance and local fiducial states with even fermionic parity only.

Following the usual approach to lattice gauge theories, our aim is to describe systems with fermionic matter. A single matter fermion must obey the transformation rules dictated by a suitable irreducible representation, that we label with $k_p$, of the gauge group $G$. As we will discuss in the following, however, not all the gauge groups allow for an irreducible representation $k_p$ whose fusion rules are compatible with the conservation of the fermionic parity, which must characterize the construction of the local fiducial states. The local physical states associated with a single fermion are described by the operators $\psi^{k_p \dag}_m$, which are of the form $a^{j_p\dag}_m$ and obey the transformation rules \eqref{righttr} and \eqref{lefttr} given by the representation $k_p$. Therefore, on a lattice vertex, the single-particle states are defined by the set $ \left\lbrace \psi^{k_p \dagger}_m \ket{\Omega_p}\right\rbrace $, where $\ket{\Omega_p}$ is the Fock vacuum and obviously transforms following the trivial representation of the group.

A local physical state with more than one matter fermion, instead, will transform with rules specified by the fusion algebra of the irreducible representations of the group $G$. We must observe, however, that the requirement of even fermionic parity of the operator $\mathcal{A}$ implies that, beside the gauge symmetry, there must be also a conserved $\mathbb{Z}_2$ fermionic number (of physical and virtual fermions together) already encoded in the local Gauss law for the fiducial states. With respect to this conserved $\mathbb{Z}_2$ charge, the irreducible representation $k_p$ must have an odd fermionic parity. These conditions can be fulfilled if the algebra defined by the fusion of the irreducible representations of the gauge group has a non-trivial $\mathbb{Z}_2$ grading \cite{Lie1990}: this means that the irreducible representations of $G$ can be distinguished into two non-empty subsets characterized by  even or odd parities, and the tensor products among them respect the conservation of such parity.
$SU(2)$, for example, satisfies this condition since we can associate an even fermionic parity to its integer representations and an odd fermionic parity to the half-integer ones. Therefore we can adopt, for example, $k_p=1/2$. In this case, by choosing the spin 1/2 representation for the single-particle matter states, the $\mathbb{Z}_2$ conservation of the fermionic number is automatically fulfilled independently of the truncation chosen for the virtual modes. This is related to the structure of $SU(2)$ which is a double covering of $SO(3)$ such that  $SU(2) \backsimeq \mathbb{Z}_2 \times SO(3)$.

Among the virtual modes, then, there will be, in general, a set of states, associated with an odd fermionic parity, which can be represented by a single fermion, and a set of states, associated to even fermionic parities, which require an even number of fermions for their construction. In the simplest case we can consider only two representations for the virtual states: the trivial representation, which is always associated to an even number of fermions and, for example, can be represented by either a vacuum state or a pair of fermions; and a non-trivial representation $j$, which may coincide with the representation $k_p$ adopted for the matter and is associated with an odd fermionic parity, in such a way that it can be encoded with a set of single-particle states.

In this case, the virtual states will be created by applying suitable operators $a^{\dag j}_m$ which transform following the rules \eqref{righttr} and \eqref{lefttr}. To build the fiducial state $\ket{A(\mathbf{x})}\equiv\mathcal{A}(\mathbf{x})\ket{\Omega}$ on each vertex, we combine the physical operators $\psi^{k_p\dagger}_m$ and the virtual operators of the kind $a^{j \dagger}_m$ that we label $l^{j \dagger}_m,l^{j \dagger}_m,u^{j \dagger}_m,d^{j \dagger}_m$, depending on the link. The operator $\mathcal{A}$ will be a sum of even products of these fermionic operators built to guarantee the consistency with the fusion rules of the gauge group and the
associated local Gauss law. Let us consider, in the following, the two main examples provided by the $U(1)$ and $SU(2)$ symmetries where this scheme which exploits only two representations is easily applied.  As one Abelian group and one non-Abelian group, these examples manifest the fundamental properties shown by such states and thus will suffice for the purpose of this paper.

In both cases, we will eventually arrive at truncated lattice gauge theories, once the physical gauge fields are introduced as well. The truncation scheme for the $U(1)$ case will be such that electric fields might take the values $0,\pm 1$, and for the $SU(2)$ case we will choose $k_p=1/2$ and we will allow the trivial and $j=1/2$ representations for the gauge field. In both cases, these are the simplest nontrivial truncations possible. As for the fermions, which shall be discussed first as we are dealing with globally-symmetric theories at this stage, the $U(1)$ will involve one physical mode at every vertex $\mathbf{x}$, associated with the creation operator $\psi^{\dagger}\left(\mathbf{x}\right)$. We will use staggered fermions \cite{Susskind1977}, and thus the charge operator will be
\begin{equation} \label{stagcharge}
Q\left(\mathbf{x}\right) = (-1)^{x_1+x_2} \psi^{\dagger}\left(\mathbf{x}\right)\psi\left(\mathbf{x}\right)
\end{equation}
(see \cite{Zohar2015b} for further details). To get the ``real'' lattice gauge theory fermionic degrees of freedom from these, one has to perform a particle-hole transformation on the odd sublattice \cite{Zohar2015b}.

In the $SU(2)$ case, there will be a $k_p=1/2$ fermionic doublet at every vertex; we will adopt the short-hand notation $\psi^{\dagger}_m\left(\mathbf{x}\right)$ where we will write $m=\pm$ for $m=\pm 1/2$. The charges are defined as
\begin{equation}
Q_a\left(\mathbf{x}\right) = \frac{1}{2} \psi^{\dagger}_m \left(\mathbf{x}\right)\left(\sigma_{a}\right)_{mn}\psi_n\left(\mathbf{x}\right)
\end{equation}
where $\sigma_a$ are the Pauli matrices and $a=x,y,z$ is the $SU(2)$ group (color) index, satisfying a right $SU(2)$ algebra:
\begin{equation}
\left[Q_a,Q_b\right]=i \epsilon_{abc}Q_c \,.
\end{equation}
To get the ``real'' lattice gauge theory fermionic degrees of freedom from our state, one has to perform a particle-hole transformation on the odd sublattice, of the form
\begin{equation}
\psi_m^{\dagger} \rightarrow \epsilon_{mn} \psi_n\,,
\label{epseq}
\end{equation}
with $\epsilon_{12}=-\epsilon_{21}=1$. Note that $\epsilon_{mn} \psi_n$ has the same transformation properties as $\psi_m^{\dagger}$, and that transformation \eqref{epseq} leaves the non-Abelian charges $Q_a$ invariant (thanks
to the fact that $SU(2)$ is a special unitary group).

For both $U(1)$ and $SU(2)$, one can use the same virtual modes. Each of the bonds may be occupied by two virtual fermionic modes,
$l^{\dagger}_m,r^{\dagger}_m,u^{\dagger}_m,d^{\dagger}_m$, with $m=\pm$. In the $U(1)$ case, they undergo group transformations generated by
the virtual electric fields
\begin{equation} \label{eflux}
E\left(a\right) = a^{\dagger}_+ a_+ - a^{\dagger}_- a_-
\end{equation}
 with $a=l,r,u,d$. Then,
 \begin{equation}
 e^{i E_a \phi}a^{\dagger}_{\pm} e^{-i E_a \phi} = e^{\pm i \phi}a^{\dagger}_{\pm}
 \end{equation}
 and the generator of the virtual Gauss law will be
 \begin{equation}
 G = E_r + E_u - E_l - E_d - Q\,.
 \end{equation}
In the $SU(2)$ case, the right and left transformations of $a$
will be generated by
\begin{equation}
R_a = \frac{1}{2}a^{\dagger}_m\left(\sigma_{a}\right)_{mn}a_n
\end{equation}
and
\begin{equation}
L_a = \frac{1}{2}a^{\dagger}_m\left(\sigma_{a}\right)_{nm}a_n
\end{equation}
which are consistent with Eqs. (\ref{righttr},\ref{lefttr}) and Eqs. (\ref{Lieright},\ref{Lieleft}). With these relations, we see that $a^{\dagger}_m\left|\Omega\right\rangle$ is transformed with the $j=1/2$ representation, and thanks to the fermionic statistics, $a^{\dagger}_+ a^{\dagger}_-\left|\Omega\right\rangle$ is a singlet (as well as $\left|\Omega\right\rangle$ of course).

The Gauss law will then be generated by the operators
\begin{equation}
G_a = L_a\left(r\right) + L_a\left(u\right) - R_a\left(l\right) - R_a\left(d\right) - Q_a
\end{equation}
satisfying a right $SU(2)$ Lie algebra.

In the $U(1)$ case, the operator creating the fiducial state is
\begin{equation} \label{AFermU1}
\mathcal{A} = \sum_{\left\{n\right\}} \alpha_{n_p n_l n_r n_u n_d} \, \delta_{Q_p+E_l+E_d,E_r+E_u} \, \psi^{\dagger n_p}
\underset{a=l,r,u,d}{\prod} a^{\dagger n^a_+}_+ a^{\dagger n^a_-}_-
\end{equation}
where the sum is over all the indices $n_i=0,1$, $Q_p = \pm n_p$ is the matter charge, whose sign is determined by the parity of the vertex (see Eq. \eqref{stagcharge}), and $E_a=n^a_+ - n^a_-$ is the virtual electric field along the link $a$ (see Eq. \eqref{eflux}), in such a way that the Kronecker delta enforces the Gauss law; this implies that the proof of the appendix applies also for fermions and we have the right symmetry.

For $SU(2)$, we write the operator creating the fiducial state as:
\begin{equation} \label{AFermSU2}
\mathcal{A} = \sum_{\left\lbrace j,m\right\rbrace} A^{j_p m_p}_{j_l m_l;j_r m_r;j_u m_u;j_d m_d}
  \underset{a=p,l,r,u,d}{\prod}\left[\delta_{j_a,0}\left(1+\tau_a a^\dag_+a^\dag_- \right)+  \delta_{j_a,1/2}  a^{\dagger}_{m_a} \right]
\end{equation}
with $A$ defined in Eq. \eqref{CGtensor}, such that the proof of the gauge invariance in \ref{app:A} applies. In this equation we adopted a slight abuse of notation: for $a=p$ the fermionic operators $a^\dag_{m_a}$ corresponds to $\psi^\dag_{m_p}$. In the previous fiducial state we introduced a set of free variational parameters $\tau_a$.
If $\tau_a \neq 0$, then the representation state $j_a=0$ is implemented twice, either with the vacuum state or with the creation of a singlet $a^\dag_+a^\dag_-$. For virtual states this implies a redundancy in obtaining the trivial representation on the links. For the physical matter state, the empty state and the singlet correspond to different physical situations in which the matter vertex is in the trivial sector, and the case $\tau_p=0$ can be interpreted as the case of infinite repulsive onsite interspecies interaction. When all the parameters $\tau_a$ are set to zero, these degeneracies in the way of obtaining the trivial representation disappear.

One can easily verify the even fermionic parity of the operators $\mathcal{A}$ for both Eq. \eqref{AFermU1} and Eq. \eqref{AFermSU2}.
First consider the $U(1)$ case. There, we have $Q_p=-E_l-E_d+E_r+E_u$, thus $-E_l-E_d+E_r+E_u-n_p=0 \mod 2$ and, given the definition of the electric fields $E$, we find that the total fermionic parity is even.
The generalization to $SU(2)$ with the virtual truncation $j=1/2$ is simple, because we get a similar result from considering the $G_z$ component of the Gauss law.

In this case, the bond projectors may be created from the vacuum by the following operators:
\begin{equation} \label{Hferm}
H\left(\mathbf{x}\right) = \prod_m\left[  1+ l^{\dagger}_m\left(\mathbf{x}+\hat{\mathbf{e}}_1\right)r^{\dagger}_m\left(\mathbf{x}\right)\right]
\end{equation}
in the horizontal case and
\begin{equation} \label{Vferm}
V\left(\mathbf{x}\right) = \prod_m\left[ 1+ u^{\dagger}_m\left(\mathbf{x}\right)d^{\dagger}_m\left(\mathbf{x}+\hat{\mathbf{e}}_2\right)\right]
\end{equation}
for the vertical bonds. Here we wrote explicitely the product over the $m$ index, no summation is intended. These operators to create the bond states are not a straightforward generalization of the previous entangled states (\ref{Hstate},\ref{Vstate}). The difference relies on the inclusion of the doubly occupied state generated by the terms $\tau_a$ in \eqref{AFermSU2}. As we discussed, these parameters correspond to a multiplicity in obtaining the trivial representation along the bonds and the projectors on the states defined by Eqs. (\ref{Hferm},\ref{Vferm}) correctly account for this multiplicity. The combination of the fiducial state \eqref{AFermSU2} and the previous link states is mapped in the construction of the previous sections in the case $\tau_a=0$.

The previous procedure has to be modified when we want to include different representations on the links, which may be associated to even or odd fermionic parities.
Let us consider the $SU(2)$ case. We keep $k_p=1/2$ for the physical matter, but we allow for more general virtual states.

All the virtual states in a half-integer representation can be built with the previous construction just by introducing further operators $a^{j\dag}_m$ that define $2j+1$ single-particle states that must be adopted to represent the group states of the related irreducible representation. Therefore, for these half-integer representations it is convenient to introduce the further constraint of having only one fermion per link, therefore $\sum_m n^a_m = 1$. This easily extends the construction of the fiducial state \eqref{AFermSU2} in the case $\tau_a=0$, where no redundancies in the definition of the physical link states are present. In principle, we can introduce also additional components of the fiducial state with a larger occupation number for the virtual half-integer representation states with $j_a>1/2$. Such virtual states with higher occupation numbers will in general be associated to the representations obtained by the fusions rules of the $j_a$ representations. This must be handled with the introduction of additional redundancy parameters which generalize $\tau_a$ and are related to the possible outcomes of the fusion of the single-particle states in the $j_a$ representation. An even occupation number of half-integer $j_a$ states, for example, will correspond to an integer representation (in the same way the singlet $a^\dag_+a^\dag_-$ is associated to the trivial representation).

For integer representations, indeed, the associated fermionic parity must be even to be consistent with the $SU(2)$ fusion rules. If we consider the integer spin $j$ representation, the $2j+1$ group states can be conveniently represented as the set of states with $\sum_m n^a_m = 2j$ fermions, or, equivalently, with the set of states representing single holes built through the operators $a^j_m$ applied to the fully occupied state. This implies only minor corrections to the previous constructions related to the different transformation laws of $a^j_m$ and the different projectors required to link the fiducial states.

In more detail, we need to redefine the operators acting on the virtual links and associated to the integer representations of $SU(2)$. These operators must be substituted by, first, fully populating the states associated to the integer representation $j$ on the virtual link acting with the operator $\prod_m a^{j\dag}_m$ on the vacuum state. Hence we define the virtual link state $\prod_m a^{j\dag}_m \ket{\Omega}$ which is an $SU(2)$ singlet composed by $2j+1$ fermions thanks to the fermionic statistics \cite{Zohar2015}. After that, we introduce the following operators for the integer representations:
\begin{equation} \label{bdef}
b^{j\dagger}_m = (-1)^m a^{j}_{-m}
\end{equation}
which fulfill the transformation properties of $a^{\dagger j}_m$. The operator $b^{j\dagger}_m$, acting on the fully occupied state for the $j$ representation, creates a hole which is indeed a many body-state composed by an even number of fermions. Therefore, by reformulating all the operators ($A,H,V$) in terms of $b^{j\dagger}_m$, their fermionic parity is guaranteed to remain even.

As an example, let us consider an $SU(2)$ invariant state corresponding to a truncation $j_{\rm max}=1$. For the sake of simplicity we specialize to the case without redundancies (thus with $\tau_a=0$ for each virtual $a$). The local fiducial state is obtained by the operator:
\begin{multline} \label{AFermSU2b}
 \mathcal{A} = \sum_{\left\lbrace j,m\right\rbrace} A^{j_p m_p}_{j_l m_l;j_r m_r;j_u m_u;j_d m_d} \left[\delta_{j_p,0}\left(1+\tau_p \psi^\dag_+\psi^\dag_- \right)+  \delta_{j_p,1/2}  \psi^{\dagger}_{m_p} \right] \times \\
  \underset{a=l,r,u,d}{\prod}\left[\delta_{j_a,0} +  \delta_{j_a,1/2}  \left( a^{j_a=1/2}_{m_a}\right)^\dag  + \delta_{j_a,1} \left( b^{j_a=1}_{m_a}\right)^\dag\prod_{m'} \left(a^{j_a=1}_{m'}\right)^\dag\right].
\end{multline}
and the local fiducial state results $\ket{A}=\mathcal{A}\ket{\Omega}$. This  fiducial state can be easily generalized through the introduction of further operators $a^{j}$ and $b^j$ for half-integer and integer representations respectively. More complicated states with redundancies in the generation of the virtual representation states may be obtained by relaxing the constraints on the number of particles in each representation species.

Concerning finite gauge groups, it should be possible to generalize such construction, based on even and odd irreducible representations, to all the double point groups (see, for example, \cite{Damus1984,Dresselhaus2008}), which constitute the finite subgroups of $SU(2)$ with the required $\mathbb{Z}_2$ grading of their irreducible representations.

\subsection{Gauging the states}
Analogously to the previous sections,  we could introduce $B$ tensor-operators for the links  for the fermionic PEPS as well, in order to include the degrees of freedom of the gauge fields and lift the global gauge symmetry to a local symmetry. In this case the operator $B$ would mix fermionic operators for the virtual fermionic modes and non-fermionic physical states describing the gauge bosons.
As shown before, however, the $B$ tensors can be unified with the $A$ tensors into a new tensor $C$; $B$ tensors constitute indeed a good starting point for a constructive explanation of the PEPS, but hereafter we will focus instead on the $C$ tensors, which allow for a more compact construction of the PEPS with local symmetries on the square lattice. The $C$ tensors, ideally corresponding to a suitable concatenation of an $A$ tensor with two $B$ tensors, will again correspond to operators in the fermionic case and will involve the matter fermions on a lattice vertex and the relates ``top'' and ``side'' bosonic degrees of freedom, analogously to Fig. \ref{fig3}.

The virtual degrees of freedom, as well as the physical ones of the matter, must be transformed to fermions as described above, in the globally-symmetric case. One can then define $C$ as a ``stator'' \cite{Reznik2002}, which is a combination of a state and an operator: this will be a state in the bosonic Hilbert space (which admits a tensor product structure, and is connected via a tensor product to the total fermionic Hilbert space) and an operator in the fermionic Hilbert space (whose problems with the tensor product structure we have already explained). However, one may avoid the stator description, remembering that a stator is obtained from an operator acting on two Hilbert spaces, after it acts on a state in only one of the spaces. Thus, we can define $C$ as an operator acting on both the bosonic and fermionic vacuum, within a Fock space composed of fermionic and non-fermionic modes.

We will choose as the bosonic vacuum the state which has electric field zero everywhere (the singlet representation for the $SU(2)$ case). Thanks to the ability to describe the bosonic Hilbert space as a tensor product, we consider each link separately. In the $U(1)$ truncated case that we analyzed in the previous section, each such local Hilbert space is three dimensional. In the basis
$\left\{\left|1\right\rangle,\left|0\right\rangle,\left|-1\right\rangle\right\}$, we define the operators
\begin{equation}
\Sigma_z = \left(
             \begin{array}{ccc}
               1 & 0 & 0 \\
               0 & 0 & 0 \\
               0 & 0 & -1 \\
             \end{array}
           \right)
\end{equation}
for the electric field, raised by
\begin{equation}
\Sigma_+ = \left(
             \begin{array}{ccc}
               0 & 1 & 0 \\
               0 & 0 & 1 \\
               0 & 0 & 0 \\
             \end{array}
           \right)
\end{equation}
and lowered by $\Sigma_- = \Sigma_+^{\dagger}$.

With these operators, $\left|0\right\rangle = \left|\Omega_b\right\rangle$, and $\left|\pm 1\right\rangle =  \Sigma_{\pm}\left|\Omega_b\right\rangle$, where $\left|\Omega_b\right\rangle$ stands for the local bosonic vacuum. Thus, in the definition of $C$ as an operator, one should include, instead of bosonic states for the $t,s$ links, the appropriate operators which will act on the physical vacuum.

 In particular this is obtained by substituting the operators $r_\pm^\dag$ and $u_\pm^\dag$ in \eqref{AFermU1} with the products $r_\pm^\dag \Sigma^s_\pm$ and $u_\pm^\dag \Sigma^t_\pm$ respectively, where the superscripts $s$ and $t$ label the side and top bosonic spaces over which the $\Sigma$ operators are acting (see \cite{Zohar2015b}). This is consistent with the previous truncation scheme. If we increase the bosonic Hilbert space in order to include $2\ell+1$ possible values of the electric field, then a larger number of virtual modes along the links must be considered and they must be suitably coupled with the $2\ell$ matrices corresponding to adequate powers of the raising and lowering operators $\Sigma$ defined in this enlarged Hilbert space.

As discussed in Sec. \ref{sec:math}, for non-Abelian groups like $SU(2)$, the inclusion of the physical link states in the fermionic PEPS is more elaborate. In this case we impose a vacuum defined by $\left|000\right\rangle = \left|\Omega_b\right\rangle$ and we define the gauge field states in the representation basis by exploiting the operator $U^j_{mn}$ and following Eq. \eqref{gaugestates}. The truncation is reflected in a truncation of the representations entering in Eq. \eqref{Udef} (see also \cite{Zohar2015}).

In particular we obtain $\left|\frac{1}{2} m n\right\rangle = \sqrt{2}U^{1/2}_{mn}\left|\Omega_b\right\rangle$ with $U^{1/2}_{mn}$ describing the truncated  group element (rotation matrix) in the representation $j=1/2$. In this case $U^{1/2}$ is a $2 \times 2$ matrix of operators acting in a 5-dimensional Hilbert space, but it is applied to the vacuum state $\ket{\Omega_b}=\ket{000}$, thus simplifying the previous expression.

For the simple truncation based on the representations $j_a=0,1/2$ only, we can obtain the correct operator $C$ by substituting in Eq. \eqref{AFermSU2} $r_m^\dag$ and $u_m^\dag$ with the products $U^{1/2}_{mn} r_n^\dag$ and $U^{1/2}_{mn} u_n^\dag$ respectively. The extension to larger truncations require a more elaborate approach in which further operators $U^j_{mn}$ are introduced and coupled with the correct fermionic operators defining the virtual states. In the case provided by Eqs. \eqref{bdef} and \eqref{AFermSU2b}, this amounts to the substitution $b^{j\dag}_m \to U^{j}_{mn} b^{j\dag}_{n}$ for the ``right'' and ``up'' links.

\section{Summary}
In this work, we have reviewed (globally) symmetric PEPS and have shown how to lift the symmetry to be local. This allows writing PEPS for lattice gauge theories, based on any gauge group $G$ which is either a compact Lie or a finite group. One can also work with truncated gauge field Hilbert spaces, which are suitable for classical simulations, following the truncation scheme of \cite{Zohar2015}.

We have described a general gauging procedure, in which, once the vertex tensors $A$ satisfy the symmetry condition (\ref{eqsym1}) and the gauge tensors $B$ satisfy (\ref{eqsym2}), along with properly constructed entangled states on the bonds, a local gauge theory is guaranteed as a direct consequence of the globally symmetric one. We also presented a detailed example, based on Clebsch-Gordan coefficients, for the $A$ tensor, which then completely parameterizes the PEPS (up to bond dimensions and variational parameters).

One can also turn to a fermionic PEPS description, which allows the inclusion of fermionic matter as in high energy physics, as was done in \cite{Zohar2015b} for $U(1)$ with Gaussian states. In this paper, we showed examples for the construction of such states, for the lattice gauge theories of the groups $U(1)$ and $SU(2)$ with given truncations.

Although our discussion in this paper has been limited to the $2+1$ dimensional case, one should note that it is easily generalizable to any $d+1$ dimensional scenario in a straightforward manner.

\section*{Acknowledgements.}

The authors would like to thank Thorsten B. Wahl and J. Ignacio Cirac, who have significantly contributed to this work while collaborating with us on PEPS for lattice gauge theories. We are also thankful to L. Tagliacozzo and H.H. Tu for helpful discussions and suggestions.
E.Z. would like to thank the Alexander von Humboldt fundation, for supporting this project through its fellowship for postdoctoral researchers. This work was supported by the EU grant SIQS.

\appendix
\section{Gauge invariance from the local fiducial states}
\label{app:A}

In this Appendix we discuss the gauge invariance of the fiducial states originating from the expression of the vertex tensor $A$ in Eq. \eqref{CGtensor}. In particular, we show that plugging the tensor (\ref{CGtensor}) into (\ref{phindef}) satisfies the Gauss law (\ref{eqsym1}):
\begin{multline}
 \left|A\right\rangle =
\underset{\left\{j\right\}}{\sum}\alpha^{j_p j_1 j_2}_{j_l j_r j_u j_d}
\left\langle j_l m_l j_d m_d | j_1 m_1 \right\rangle
\left\langle j_1 m_1  j_p m_p |  j_2 m_2 \right\rangle \left\langle j_2 m_2 | j_r m_r j_u m_u \right\rangle \times \\
\left|j_p m_p;j_l m_l;j_r m_r;j_u m_u;j_d m_d\right\rangle \equiv
\underset{\left\{j\right\}}{\sum} A^{j_p m_p}_{j_l m_l;j_r m_r;j_u m_u;j_d m_d} \left|j_p m_p;j_l m_l;j_r m_r;j_u m_u;j_d m_d\right\rangle.
\end{multline}
To this purpose, let us apply the transformations on the fiducial state $\left|A\right\rangle$:
\begin{multline}
\ket{A' \,} \equiv \Theta_g \left|A\right\rangle =
\Theta_g^p  \tilde \varTheta^{r \dagger}_g \tilde \varTheta^{u \dagger}_g \varTheta^{l}_g \varTheta^{d}_g \left|A\right\rangle =
\underset{\left\{j\right\}}{\sum}A^{j_p m_p}_{j_l m_l;j_r m_r;j_u m_u;j_d m_d} \times \\
D^{j_p}_{m'_p m_p} \left(g\right) D^{j_l}_{m'_l m_l} \left(g\right) D^{j_d}_{m'_d m_d} \left(g\right)
\overline{D^{j_r}_{m'_r m_r}} \left(g\right) \overline{D^{j_u}_{m'_u m_u}} \left(g\right)
\left| j_p m'_p j_l m'_l j_r m'_r j_u m'_u j_d m'_d \right\rangle,
\end{multline}
using the Clebsch-Gordan series \cite{Zohar2015}, we obtain
\begin{equation}
D^{j_l}_{m'_l m_l} \left(g\right) D^{j_d}_{m'_d m_d} \left(g\right) = \underset{J}{\sum}D^{J}_{M'M}\left(g\right)
\left\langle j_l m'_l j_d m'_d | JM' \right\rangle
\left\langle JM | j_l m_l j_d m_d \right\rangle
\end{equation}
and
\begin{equation}
\overline{D^{j_r}_{m'_r m_r}} \left(g\right) \overline{D^{j_u}_{m'_u m_u}} \left(g\right) = \underset{K}{\sum}\overline{D^{J}_{N'N}}\left(g\right)
\left\langle KN' | j_r m'_r j_u m'_u \right\rangle
\left\langle j_r m_r j_u m_u | KN \right\rangle.
\end{equation}
Using completeness within the representations, we get:
\begin{equation}
 \left\langle J M | j_l m_l j_d m_d \right\rangle \left\langle j_l m_l j_d m_d | j_1 m_1 \right\rangle = \delta_{J,j_1}\delta_{M,m_1}
\end{equation}
as well as
\begin{equation}
\left\langle j_2 m_2 |  j_r m_r j_u m_u \right\rangle \left\langle j_r m_r j_u m_u | KN \right\rangle = \delta_{j_2,K}\delta_{m_2,N}.
\end{equation}
Thus,
\begin{multline}
\ket{A'\,} =
 \underset{\left\{j\right\}}{\sum}\alpha^{\left\{j\right\}}
D^{j_1}_{M'm_1}\left(g\right)   \overline{D^{j_2}_{N'm_2}}\left(g\right)   D^{j_p}_{m'_p m_p}\left(g\right) \times \\
\left\langle j_l m'_l j_d m'_d | j_1 M' \right\rangle \left\langle j_1 m_1 j_p m_p | j_2 m_2 \right\rangle
\left\langle j_2 N' | j_r m'_r j_u m'_u \right\rangle \left| j_p m'_p j_l m'_l j_r m'_r j_u m'_u j_d m'_d \right\rangle.
\end{multline}
We use the Clebsch-Gordan series and completeness again, to obtain
\begin{equation}
D^{j_1}_{M' m_1} \left(g\right) D^{j_p}_{m'_p m_p} \left(g\right) = \underset{J}{\sum}D^{J}_{I'I}\left(g\right)
\left\langle j_1 M' j_p m'_p | JI' \right\rangle
\left\langle JI | j_1 m_1 j_p m_p \right\rangle
\end{equation}
and
\begin{equation}
 \left\langle J I | j_1 m_1 j_p m_p \right\rangle \left\langle j_1 m_1 j_p m_p | j_2 m_2 \right\rangle = \delta_{J,j_2}\delta_{I,m_2}
\end{equation}
resulting in
\begin{multline}
\ket{A'\,} =
 \underset{\left\{j\right\}}{\sum}\alpha^{\left\{j\right\}}
D^{j_2}_{I'm_2}\left(g\right)   \overline{D^{j_2}_{N'm_2}}\left(g\right) \times \\
\left\langle j_l m'_l j_d m'_d | j_1 M' \right\rangle \left\langle j_1 M' j_p m'_p | j_2 I' \right\rangle
\left\langle j_2 N' | j_r m'_r j_u m'_u \right\rangle \left| j_p m'_p j_l m'_l j_r m'_r j_u m'_u j_d m'_d \right\rangle.
\end{multline}
Finally, thanks to the unitarity of the Wigner matrices,
\begin{equation}
D^{j_2}_{I'm_2}\left(g\right) \overline{D^{j_2}_{N'm_2}}\left(g\right) = D^{j_2}_{I'm_2}\left(g\right) D^{j_2}_{m_2 N'}\left(g^{-1}\right) = \delta_{I'N'}
\end{equation}
and altogether we obtain
\begin{equation}
\ket{A' \,} =  \left|A\right\rangle
\end{equation}
which completes the proof.

Note that for the proof one does not have to sum on the representation indices - every summand respects the symmetry, and thus the representations are the indices of the tensor.

\newpage

\section*{References}
\bibliography{ref}

\end{document}